\documentclass[journal]{IEEEtran}
\usepackage{cite}

\ifCLASSINFOpdf
  \usepackage[pdftex]{graphicx}
  \graphicspath{{pdf/}{png/}{jpeg/}}
\else
  \usepackage[dvips]{graphicx}
  \graphicspath{{eps/}}
\fi
\usepackage[cmex10]{amsmath}
\usepackage[caption=false,font=small]{subfig} 
\usepackage{placeins} 

\usepackage{epsfig}

\usepackage{pgfplots}
\pgfplotsset{compat=newest}
\pgfplotsset{plot coordinates/math parser=false}
\newlength\figureheight
\newlength\figurewidth 

\usepackage{bm} 										
\newcommand{\uvec}[1]{\boldsymbol{\hat{\textbf{#1}}}}   
\renewcommand{\vec}[1]{\bm{#1}} 						
  
\usepackage{multirow}

\usepackage{graphicx}
\usepackage{color}
\usepackage{placeins}
\usepackage{float}
\usepackage{hyperref}
\usepackage{tabularx,colortbl}

\begin{document}
%
\title{
%
On the Accuracy of Equivalent \\Antenna Representations

}
%
%
%

%

\author{Johan~Malmstr\"{o}m, 
        Henrik~Holter, 
        and~B.~L.~G.~Jonsson
\thanks{}
\thanks{J.~Malmstr\"{o}m and B.~L.~G.~Jonsson are with the School of Electrical Engineering,
KTH Royal Institute of Technology, Stockholm, Sweden.}
\thanks{J.~Malmstr\"{o}m and H.~Holter are with Saab Surveillance, Stockholm, Sweden. The work 
was supported by Saab Surveillance.}

	}

\markboth{Malmstr\"{o}m \MakeLowercase{\textit{et al.}}: On the Accuracy of Equivalent Antenna Representations}{}            	
%

\maketitle

\begin{abstract}
The accuracy of two equivalent antenna representations, near-field sources and far-field sources, are evaluated 
for an antenna installed on a simplified platform in a series of case studies using different configurations of equivalent antenna representations. 
{The accuracy is evaluated in terms of installed far-fields and surface currents on the platform.}
%
The results {show large variations between configurations. The  root-mean-square installed far-field error is $4.4\,$\% {for} the most accurate equivalent representation.}
When using far-field sources, {the design parameters have a large influence of the achieved accuracy.} 
{There is also a varying accuracy depending on the type of numerical method used.}
Based on the results, some recommendations on the choice of {sub-domain for the equivalent antenna representation are given}. 
%
{In industrial antenna applications, the accuracy in determining e.g. installed far-fields and antenna isolation on large platforms are critical. 
Equivalent representations can reduce the fine-detail complexity of antennas and thus give an efficient numerical descriptions to be used in large-scale simulations. The results in this paper can be used as a guideline by antenna designers or system engineers when using equivalent sources.}

\end{abstract}


%
\IEEEpeerreviewmaketitle

\vspace{0pt}
%
\section{Introduction}
\label{sec:Introduction}



\IEEEPARstart{T}{he increasing} accuracy of full-wave large-scale simulation of complex electromagnetic (EM) problems has changed the industrial design-process of radio frequency (RF) systems. The construction of scaled models for measurements has widely been replaced by simulations. 
This work flow saves both time and money as compared to building prototypes for measurements \cite{Macnamara2010}. 
{For simulations to be trustworthy, in particular in industrial design processes, an a-priori predictable accuracy is essential.}

{The reason for the large interest in simulations within the electromagnetic community is that} electromagnetic scattering problems are rarely possible to solve analytically. Problems involving realistic antennas are therefore usually solved numerically.
For small-scale systems, a detailed model of the physical antenna serves as a good model for accurately determining associated fields, impedances, and currents {with} numerical methods. This is documented by e.g. the annual benchmarks given by the EurAAP working group on software \cite{EurAAP_WG4}, {where different simulation methods and measurements are compared} \cite{Vandenbosch2016}. The simulations have shown an increasing agreement with measurements over the last few years \cite{Vandenbosch2016a}. 
For electrically large platforms, it is problematic to include models of the antenna and the complete platform in the same simulation, since the computational time and memory requirement grow with the electrical size \cite{RylanderEtAl2013}. 

{Simulations of electrically large platforms in combination with complex antennas}, can lead to extreme computational time and memory requirement. In these cases, a less complex antenna model can significantly reduce the overall complexity of the simulation. Using an equivalent representation of the antenna is one way to reduce the complexity of the antenna model. 
{An equivalent antenna representation also opens the possibility to utilize different numerical methods in  different parts of the simulation domain, sometimes referred to as a {hybrid method}.} 

On large platforms, special techniques can be used to avoid limitations of platform size from memory requirements. {One commonly used technique is  asymptotic methods, e.g. physical optics (PO) or geometrical optics (GO), which are high-frequency approximations. Another technique} is domain decomposition that split a large simulation into several smaller sub-problems that are solved in parallel \cite{ToselliWidlund2005}. This has been used in e.g. \cite{Zhao2008, BeckerHansenhybrid, BarkaCaudrillier2007, Foged2015}. A drawback with domain decomposition is that the complexity is increased as compared to a full-domain simulation, since information has to be exchanged over the interface between adjacent sub-domains. 

One way of decomposing a large EM problem is to analyze the antennas in isolation, and imprint the results in the platform model. In a basic domain decomposition, multiple scattering effects between structures in different domains are neglected, which implies that there is no need for iterations over the interface between adjacent sub-domains. In that case, the sub-domain with the antenna serves as an equivalent representation of the antenna, in the same way as in this work.
When installing antennas on a platform, the antennas are often placed so that the  scattering from the platform is minimized. Therefore, neglecting multiple scattering is often a minor approximation in antenna placement studies.

Commercial software for computational electromagnetics (CEM), see e.g. \cite{CSTMWS2016, HFSS2016, FEKO2016, COMSOL2016}, and recommendations for CEM has been thoroughly discussed in the literature see e.g. \cite{Weiland2008}. From an industrial perspective it is of major interest to get a highly accurate solution, but equally important to get a predictable accuracy. 
{Simulation software documentation can give  designers valuable rule-of-thumb recommendations} about preferred equivalent model configurations, but the recommendations seldom give any information about the expected accuracy when using the equivalent representation.
Unfortunately, it is not at all obvious which configuration of the equivalent {antenna model} that is most accurate. 

Source modeling in electromagnetics has been of scientific interest for a long time, where major contributions have been made by e.g. Hall\'{e}n, King and Harrington. Some of these works are reviewed in {e.g.} \cite{Balanis1992}.
These classical works focus mainly on antenna excitations for a specific type of antenna, e.g. wire antennas. Here, we do not model the feeding, but rather use an equivalent model for the entire antenna.
The two here studied equivalent representations, near-fields sources and far-field sources, are general in the sense that they can be used for any type of antenna.
The underlying theories for these two representations, {the equivalence principle and far-field pattern generation}, are both described in literature; see e.g. \cite{Balanis2005} and the references within. 
They are also implemented in several commercial software, see e.g. \cite{CSTMWS2016, FEKO2016}. {Results from using equivalent representations are described in several articles, e.g. \cite{Foged2014, Foged2015, Wang2013, Li2016, Payet2014}, where near-field sources are used}, and e.g. \cite{Tanjong2010} that shows examples when using far-field sources. 

The results in this paper add to the practical knowledge by examining the accuracy of equivalent antenna representations, for different parameter configurations and different numerical methods.
In addition to the installed far-field, which is one of the most important characteristic of an installed antenna, we also evaluate the induced surface current on the platform, which is important in both EMC applications and in antenna {placement studies} \cite{Macnamara2010}.
{The results indicates what order of magnitudes of errors that  equivalent antenna models introduce.} The results also serve as a guideline for an antenna designer in order to choose the most accurate configuration when using equivalent representations of antennas. 

The paper consists of the following sections. Section~\ref{sec:Theory} describe the theory used in the article and  Section~\ref{sec:Method} the used methods and models. The results are presented in Section~\ref{sec:Results}. We evaluate the accuracy of near-field sources in Section~\ref{sec:Results1}, far-field sources in Section~\ref{sec:Results2}, and their combination with different numerical methods in Section~\ref{sec:Results3}. The paper ends with a discussion and conclusions in Section~\ref{sec:Conclusions}.

\vspace{0pt}
%
\vspace{-1mm}
\section{Equivalence Theory} 
\label{sec:Theory}

Two types of equivalent antenna representations are evaluated in this paper; near-field sources and far-field sources. 
Both representations are applicable to any kind of antenna or other radiating structure. 


\subsection{Near-field Sources}
The equivalence principle, introduced by Love \cite{Love1901} and refined by Schelkunoff \cite{Schelkunoff1936}, implies that any radiating structure can be represented by electric $\vec{J}$ and magnetic $\vec{M}$ surface currents on an fictitious surface $\Gamma_e$ enclosing the radiating structure \cite{StutzmanThiele1998}. 
If the material outside $\Gamma_e$ is homogeneous, 
it can be deduced that $\vec{J}$ and $\vec{M}$ 
reproduce exactly the same electric and magnetic fields outside $\Gamma_e$ as the original antenna, while the fields inside 
$\Gamma_e$ are zero \cite{StutzmanThiele1998,Balanis1997}. 
A near-field source (NFS) is a set of surface currents $\vec{J}$, $\vec{M}$ on the surface $\Gamma_e$, a \textit{Huygens' surface}. With the presence of a platform, the homogeneous requirement is not fulfilled and the near-field source only  reproduces the original fields approximately. 


The surface currents $\vec{J}$ and $\vec{M}$, which we hereafter refer to as \emph{currents}, are directly related to the discontinuity of the tangential electric and magnetic fields on the surface $\Gamma_e$. 
Since the currents $\vec{J}$, $\vec{M}$ produce zero fields inside the surface $\Gamma_e$, the currents on $\Gamma_e$ can be written as \cite{StutzmanThiele1998, Jackson1998}
\begin{eqnarray}
\label{eq:eq_surf2a}
\vec{J} &=& \uvec{n} \times \vec{H},\\
\label{eq:eq_surf2b}
\vec{M} &=& - \uvec{n} \times \vec{E},
\end{eqnarray}
where $\vec{E}$ and $\vec{H}$ are the electric and magnetic fields on $\Gamma_e$ {and $\uvec{n}$ is the outward pointing normal to the surface $\Gamma_e$}. 

It can be beneficial to use a near-field representation of an antenna instead of a model of the physical antenna. One reason is if the number of mesh cells in the model including the platform decreases. Also, it opens up for using different numerical methods when analyzing the antenna and the platform.
 
{When an antenna is installed on a platform}, the surrounding material is non-homogeneous; the platform is typically made of conducting or dielectric material that is surrounded by e.g. air or vacuum. 
Using a near-field source to represent the antenna in a situation when the homogeneous condition is violated introduces an error {because the surface currents are not correctly described}. 
Hence, using a near-field source under this common circumstance is an approximation. {The error from this approximation is effected by the choice of sub-domain, and is evaluated in Section~\ref{sec:Results1}}.
%
The generation of a near-field source is described in Section~\ref{sec:workflow_NFS}.

\subsection{Far-field Sources}
Antenna radiation patterns are commonly used to characterize antennas. They describes magnitude, phase and polarization of the propagating waves generated by the antenna, for directions $\varphi,\theta$, and a fixed distance $r$. 
The \textit{far-field radiation pattern} is the leading order behavior as $r\rightarrow \infty$ \cite{StutzmanThiele1998}. 


A far-field radiation pattern can be used as an equivalent source in electromagnetic simulations. We will then refer to it as a far-field source (FFS). For reciprocal antennas, it also describes the receiving characteristics of the antenna. 
The far-field source is characterized by its far-field radiation pattern imposed as an infinitesimally small source at a given position. This position together with the approximation of the platform when generating the far-field source, are the design parameters for far-field sources. Their impact on the accuracy of the solution is non-trivial, and is examined in Section \ref{sec:Results2}. 

Far-field sources are, compared to near-field sources, more a primitive representation, and are not expected to perform as good. On the contrary, they provide very efficient representations {in numerical methods, from an implementation point of view}. 
The error introduced by the far-field approximation is small when the far-field source is far from surrounding structure, e.g. a horn antenna feeding a parabolic reflector, see e.g. \cite{Yaghjian1984}.
However, far-field sources {have} be used also near structures, see e.g. \cite{Tanjong2010}. 

{Multipole expansion is a procedure to increase the accuracy of a far-field approximation close to the source point, i.e. in the near field region \cite{Jackson1998}. However, with a far-field source installed on a platform, multipole expansion cannot be used, since it requires that there is no charge or current carrying structure within a certain distance from the center point of the expansion \cite{Jackson1998}. Multipole expansion is therefore not further considered in this paper.} 


Using a far-field source, its emitted radiation will locally be a plane wave, irrespective of the true distance between the far-field source and an evaluation point. 
{A far-field source can be used e.g. in an integral equation formulation, where the point source is used as a field source within the computation domain.} Implementation details of the point-shaped far-field source depend on the type of numerical method used, see e.g. \cite{Gibson2008, Makarov2002, RylanderEtAl2013} for general discussions.
The generation of a far-field source is described in Section~\ref{sec:workflow_FFS}.

%

\vspace{0pt}
%
\section{Methods} 
\label{sec:Method}

We use a series of {case studies} to investigate the accuracy of equivalent antenna representations for different {configurations}. For electromagnetic problems, where analytic solutions can rarely be found, the case study is an effective tool to compare configurations. The aim is to determine how equivalent antenna representations affect the accuracy of fields and currents in the presence of a platform  {and to obtain an understanding of the relation between configuration of the equivalent antenna representation and accuracy}. 
%
\subsection{Geometrical Model}
We choose here a platform geometry $G$ and an antenna $A$. 

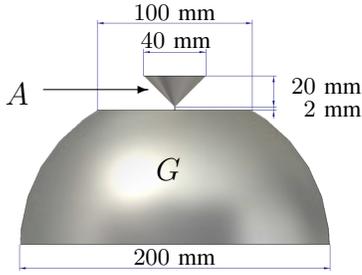
\begin{figure}[!t]
	\centering
		\setlength{\unitlength}{1mm}
		\begin{picture}(80, 35)(-10, 0)
			\put(5,0){
			    	\includegraphics[width=45mm, trim=70mm 18mm 160mm 13mm, clip]{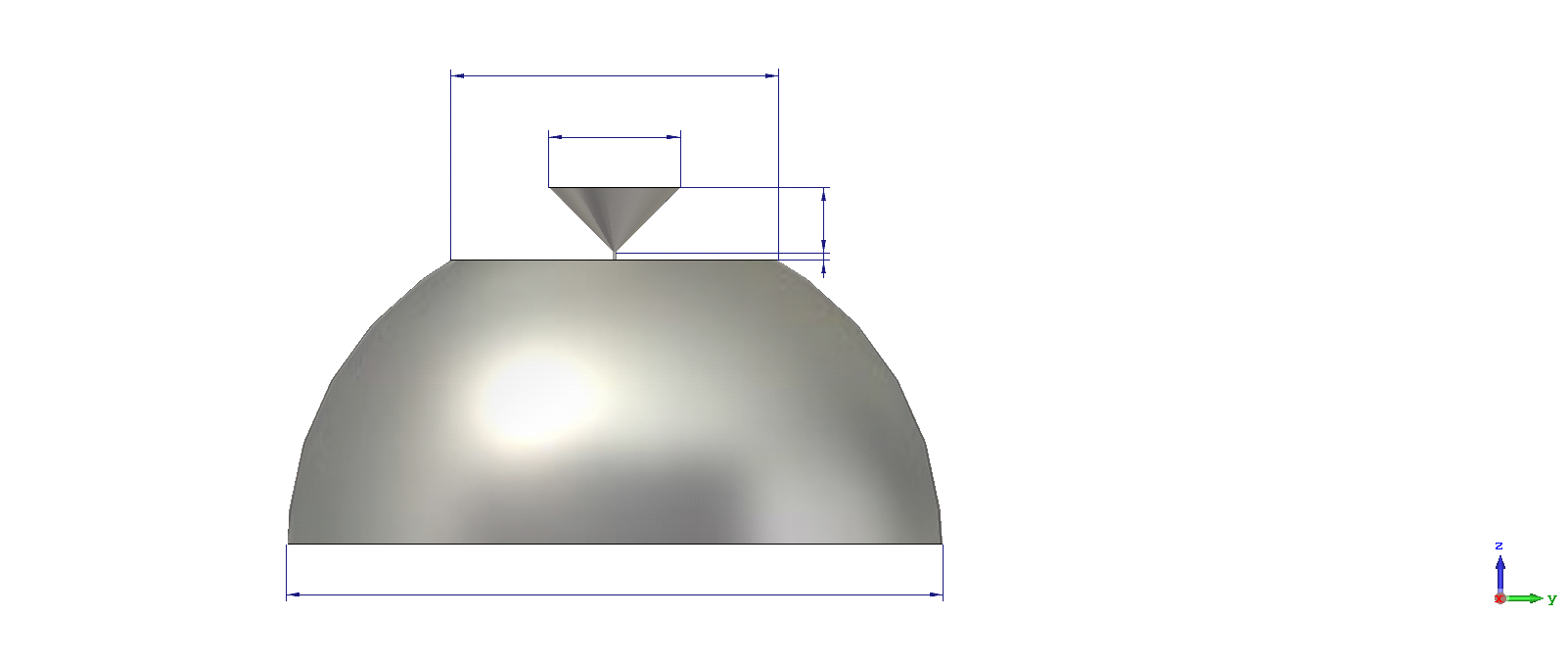}
		   	}
		   	\small
			\put(23,00.7){$200$~mm}
			\put(23,33.2){$100$~mm}			
 			\put(24,29.6){$40$~mm}
 			\put(44,23.4){$20$~mm}
 			\put(44,20.4){~$\,2$~mm}
 			\large 
 			\put(26,12){$G$}
 			\put(06,22){$A$}
 			\put(11,24){\vector(1,0){14}}
		\end{picture}
	\caption{The platform geometry $G$ with the mono-cone antenna $A$ placed on its flat top.}%
	\label{fig:main_structure}
\end{figure}

The platform geometry $G$ should have a simple and well defined shape, but still retaining platform specific features 
for installed antennas.
The chosen platform geometry $G$ is depicted in Fig.~\ref{fig:main_structure}. The geometry is a cut $200$~mm diameter half-sphere, where the diameter of the cut top is $100$~mm. It is perfectly electrically conducting (PEC).
The geometry $G$ is of electrically size about $7 \,\lambda$ at 10 GHz and has the following platform like properties: 

\begin{itemize}
\item Finite support of surface currents,
\item weak back-scattering from its edges to the antenna, 
\item radiating creeping waves {on} the curved surface. 
\end{itemize}

The antenna $A$ chosen for this case-study is a mono-cone antenna, also depicted in Fig.~\ref{fig:main_structure}, with $20$~mm top radius and $20$~mm height. The mono-cone antenna is mounted centrally on the flat top surface of the main geometry with a $2$~mm feed gap. It is fed with a wave-guide port via a $10$~mm long coaxial cable. A mono-cone antenna has an isolated far-field radiation pattern similar to a mono-pole but is more wideband. 


We are interested in determining fields and currents in the domain on and outside the  geometries $G\cup A$. Since the main geometry $G$ and the mono-cone antenna $A$ are both rotational symmetric, it implies that fields and currents must have the same rotational symmetry. 
It is hence sufficient to evaluate symmetric quantities on a plane $(r,\theta)$ and a fixed azimuth $\varphi$. 

\subsection{Calculation Domains}
\begin{figure}[!t] 
	\centering
		\setlength{\unitlength}{1mm}
		\begin{picture}(80, 23)(0, 1)
			\put(0,05){
			    	\includegraphics[width=78mm, trim=46mm 137mm 40mm 132mm, clip] {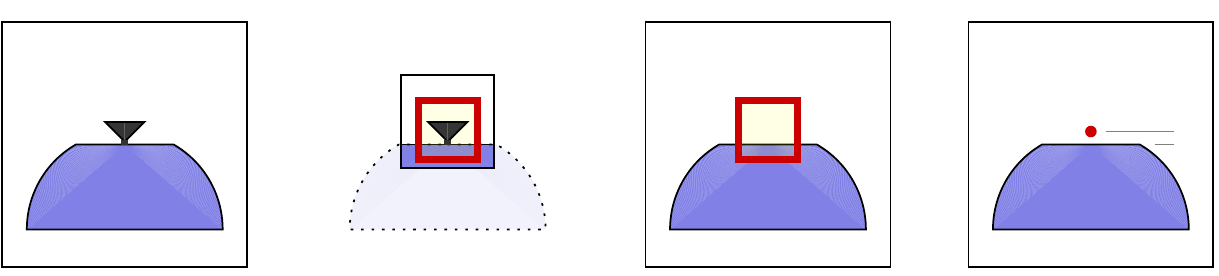}
    		}
			\put( 2,22){$D_0$}
			\put( 7,01){(a)}
			\put(28,01){(b)}		
			\put(24,18.5){$D_1$}
			\put(32,14){$\Gamma_e$}
			\put(49,01){(c)}		
			\put(44,22){$D_0$}
			\put(53.0,14){$\Gamma_e$}
			\put(70,01){(d)}		
			\put(65,22){$D_0$}
			\put(76.6,12.6){$h$}			
		\end{picture}
	\caption{Calculation domains: (a) The full domain $D_0$ is used when solving the reference case, (b) the sub-domain $D_1$ is used when generating the equivalent sources.   The domain $D_0$ is used for imprinting the equivalent sources, (c) near-field sources on the interface $\Gamma_e$ and (d) point-shaped far-field sources placed a distance $h$ above the flat platform top surface.}
	\label{fig:domains}
\end{figure}
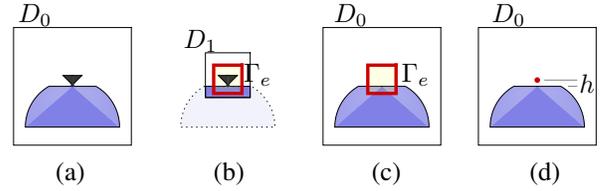

The following domain definitions, as in Fig.~\ref{fig:domains}, are used:
\begin{itemize}
\item The domain $D_0$ that contain the platform geometry $G$ and the antenna $A$. 
\item A sub-domain $D_1$ that contain the antenna and a selected part of the platform. 
{As the equivalent source is intended to improve calculation time and memory use, the sub-domain $D_1$ should be much smaller than the domain $D_0$.}
\item A Huygens' surface  $\Gamma_e$, i.e. a fictive surface enclosing the antenna. The Huygens' surface $\Gamma_e$ is completely included in the sub-domain $D_1$. 
\end{itemize}
In a platform specific problem, e.g. with aircraft or ships, the domain $D_0$ is often electrically very large. 
However, in this work, we use a comparatively small domain $D_0$. 
{This allows us to obtain a highly accurate reference solution for the fields and currents from the installed antenna, which is subsequently used as our benchmark.}
The equivalent sources can be used on electrically large structures as well, {so the methods and observations are applicable to antennas on e.g. ships or aircraft}.

An antenna induces currents on surrounding structures. Such currents also contribute to the equivalent source. By using the small domain $D_1$ to determine the equivalent sources, we implicitly ignore all  currents outside $D_1$. The larger the sub-domain $D_1$ is, the smaller the error will be, but the computational cost to determine the equivalent source increases with growing size of $D_1$.

\subsubsection{Work Flow for Using an Equivalent Near-field Source}
\label{sec:workflow_NFS}
\begin{itemize}
	\item In the sub-domain $D_1$; Calculate the tangential electric and magnetic fields, $\uvec{n}\times\vec{E}$ and $\uvec{n}\times\vec{H}$, on the surface $\Gamma_e$ generated by the antenna, see Fig.~\ref{fig:domains}(b).
	\item In the domain $D_0$; Remove the model of the physical antenna%
		\footnote{Instead of removing the antenna model, it can be replaced by a simplified model to include some of the multiple scattering effects that the model of the physical antenna model would give. 
		} %
	and imprint the equivalent currents $\vec{J} = \uvec{n}\times\vec{H}$ and $\vec{M}=-\uvec{n}\times\vec{E}$ on $\Gamma_e$, see Fig.~\ref{fig:domains}(c).
	\item Solve the problem in the domain $D_0$ using $\vec{J}$, $\vec{M}$ as equivalent sources. 
\end{itemize}
We use a rectangular box as surface $\Gamma_e$ for the equivalent currents that define the near-fields source. 

\subsubsection{Work Flow for Using an Equivalent Far-field Source}
\label{sec:workflow_FFS}
\begin{itemize}
	\item In the sub-domain $D_1$; Calculate the far-field radiation pattern generated by the antenna, see Fig.~\ref{fig:domains}(b). This is the equivalent far-field source. 
	\item Choose a height $h$ over the main geometry where the far-field source is imprinted.
	\item In the domain $D_0$; Remove the model of the physical antenna and solve the problem using the far-field source, see Fig.~\ref{fig:domains}(d).
\end{itemize}

The procedures described above in \ref{sec:workflow_NFS}--\ref{sec:workflow_FFS} is general and applies to any kind of structure, not only the platform and {the} antenna used in this case study.

\subsection{Evaluation}
\label{sec:Method_evaluation}
The results in e.g. \cite{EurAAP_WG4,Vandenbosch2016,Vandenbosch2016a}  indicate that full-wave simulation results can be used as a benchmark of antenna behavior. 
We evaluate the electric current and the installed far-field pattern by comparing simulations with equivalent antenna representations against {an} accurate reference solution using the  model of the physical antenna.

In order to get reliable estimates of the accuracy, it is crucial to have a highly accurate reference solution. We verify that the solution is accurate by solving the reference case with three different numerical methods; Finite integration technique (FIT), Method of moments (MoM), Finite element method (FEM), see e.g. \cite{TafloveHagness2000, Chew2001fast, RylanderEtAl2013}.  

When comparing results from simulations, one must keep in mind that different settings, e.g. the mesh, will affect the results. The results in this paper are from simulations where the solver settings are not the {main} limiting factor for the accuracy, but rather the representation of the antenna.

Two of the most important antenna quantities are {installed far-field patterns and isolation between antennas} \cite{Macnamara2010}. 
The installed far-field patterns are of large importance when determining the functionality of a radio frequency (RF) system, which is determined e.g. by the coverage for a sensor installed on a vehicle. 
There can be a large difference between an antennas \textit{isolated radiation pattern}, which is the radiation pattern for the antenna in free space, and its \textit{installed radiation pattern}, which is the radiation pattern when installed on a platform. Hence, as is well known, antennas must be {analyzed} within their complete environment \cite{LepvrierEtAl2014}.
The isolation between antennas is related to the risk for interference between the antennas. It is an important quantity in antenna placement studies \cite{Macnamara2010}. 
The isolation between antennas are mainly determined by direct propagating waves and surface currents, see e.g. \cite{Malmstrom2016}.
The accuracy of surface currents is thus an important contribution to the accuracy of isolation between antennas, in particular for antennas without line-of-sight.

\begin{figure}[!t] 
	\centering
		\setlength{\unitlength}{1mm}
		\begin{picture}(80, 28)(-3, 0)
			\put(4,0){
			    	\includegraphics[height=28mm, trim=185mm 37.8mm 185mm 31mm, clip] {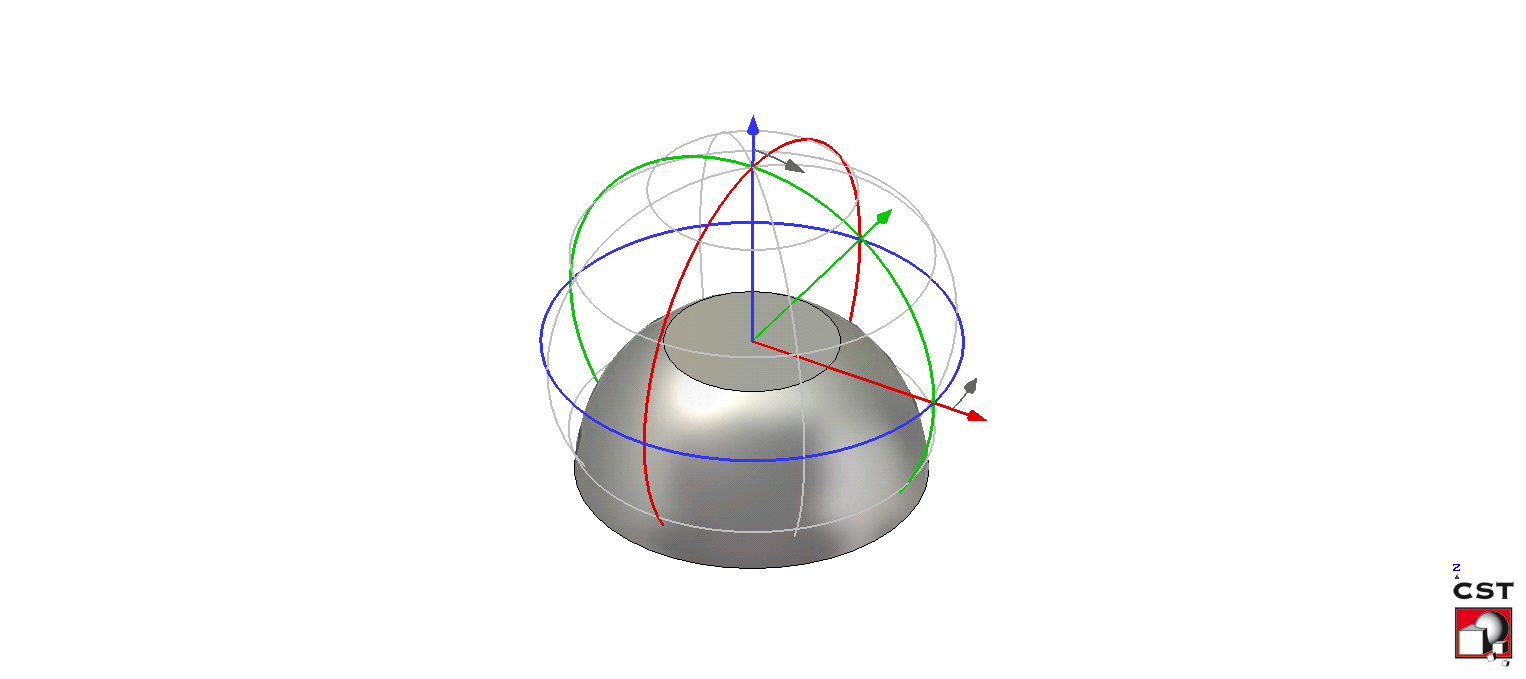}
			}
			\normalsize
			\put(31.5,6.5){$x$}
			\put(26.5,21.0){$y$}			
			\put(15.0,26.5){$z$}			
			\put(21.0,21.5){$\theta$}			
			\put(31.0,12){$\varphi$}
			\put(46,0){
			    	\includegraphics[width=30mm, trim=290mm 4mm 00mm 69mm, clip] {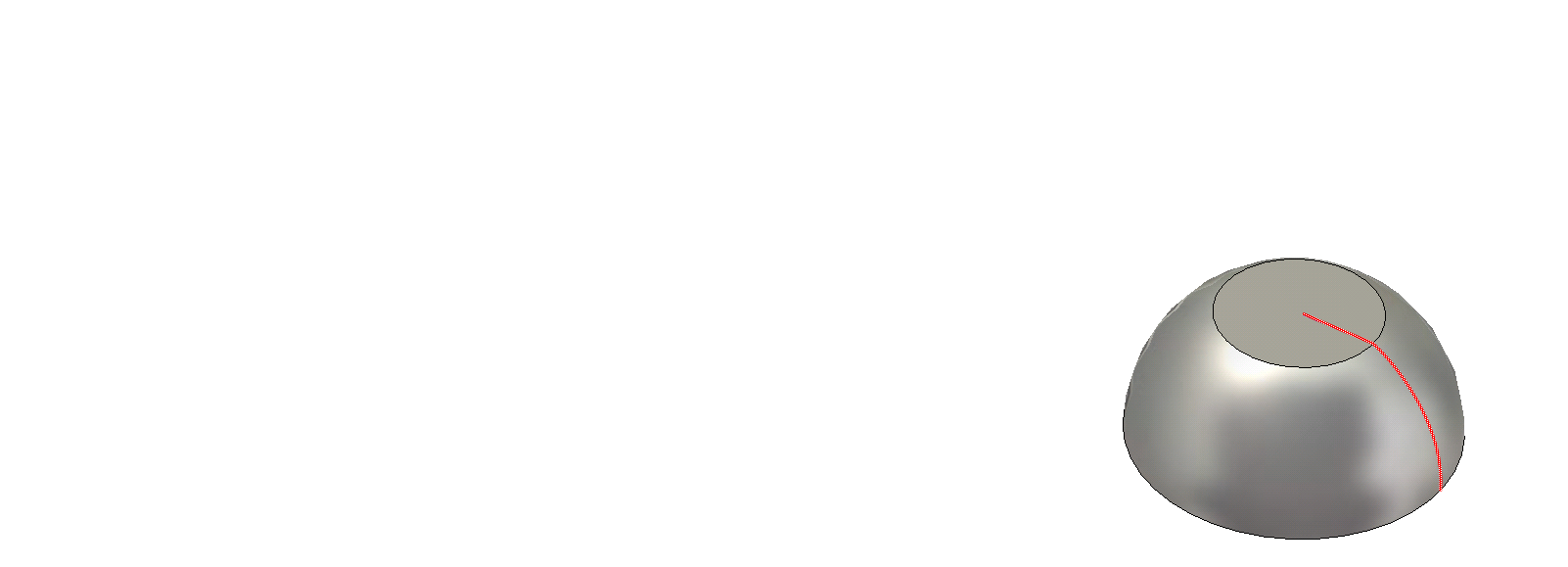}
			}
			{\small			
				\put( 2,1){(a)}
				\put(46,1){(b)}
			}
		\end{picture}
	\caption{Definition of (a) the spherical angles $\theta$ and $\varphi$, {and their relation to the Cartesian coordinates $x,y,z$,} 
	(b) the curve used for evaluation of currents. The curve is parametrized by the arc length $\ell$, where $\ell=0$ is the center of the flat top, and $\ell\approx 155$~mm is the periphery of the bottom of the geometry.
	}%
	\label{fig:main_structure_with_curve}
\end{figure}
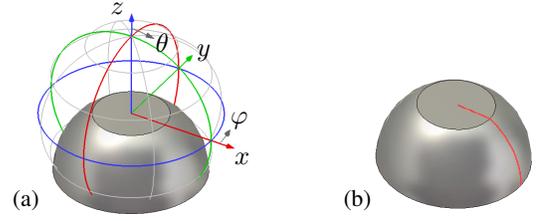

The following quantities are used for evaluating the accuracy of the results using the equivalent sources:
\begin{itemize}
\item The electric current $\vec{J}$ along the red curve in Fig.~\ref{fig:main_structure_with_curve}(b). 
\item The installed far-field $\vec{E}(\varphi,\theta)$
for all inclination angles $\theta=[\,0,180^\circ]$ and a fixed azimuth $\varphi = 90^\circ$. 
\end{itemize}

We use \eqref{eq:eq_surf2a} to evaluate the electrical current $\vec{J}$ based on the magnetic field $\vec{H}$. 
The magnetic field $\vec{H}$ can be separated into components parallel and orthogonal to the azimuthal unit vector $\uvec{e}_\varphi$,
\begin{eqnarray}
\label{eq:H_comp_phi}
H_\varphi &=& \uvec{e}_\varphi \cdot \vec{H},\\
\label{eq:H_comp_rtheta}
|H_{r,\theta}| &=& |\uvec{e}_\varphi \times \vec{H}|. 
\end{eqnarray}
The propagating component $H_\varphi$ is dominant outside the reactive region, so that we can make the approximation $\vec{H} \approx H_\varphi \uvec{e}_\varphi$. 
With this approximation, the electric current $\vec{J}$ can be calculated as
\begin{eqnarray}
\label{eq:measure_Ht}
\vec{J} \;=\; \uvec{n} \times \vec{H} 
\;\approx\; H_\varphi \, (\uvec{n} \times \uvec{e}_\varphi).
\end{eqnarray}
The rotational symmetry implies that $\uvec{n} \times \uvec{e}_\varphi = -\uvec{e}_t$, where $\uvec{e}_t$ is the outward pointing tangent unit vector along the curve in Fig.~\ref{fig:main_structure_with_curve}(b). The tangential current ${J}_t$ along the curve is
\begin{eqnarray}
\label{eq:measure_Jt}
J_t \;=\; \uvec{e}_t \cdot \vec{J} \;\approx\;  -H_\varphi.
\end{eqnarray}
%
We denote the complex tangential current from the references solution $J^\text{ref}_t(\ell)$ and when using an equivalent source $J_t(\ell)$. The scalar $\ell$ is the arc length along the curve in Fig.~\ref{fig:main_structure_with_curve}(b). 
%
We define the relative magnitude error $\delta_\text{rel} |J_t(\ell)|$ and the phase error $\angle \delta J_t(\ell)$ of the complex current as
\vspace{-1mm}
\begin{eqnarray}
\label{eq:measure_dJt_rel}
\delta_\text{rel} |J_t(\ell)| &=& 
\frac{|J_t(\ell)| - |J^\text{ref}_t(\ell)|}{|J^\text{ref}_t(\ell)|}
,\\
\label{eq:measure_dJt_phase}
\delta \angle J_t(\ell) &=&
\angle J_t(\ell) - \angle J^\text{ref}_t(\ell),
\end{eqnarray}
where $\angle(\cdot)$ denotes the argument of a complex number. 

The installed electric far-field $\vec{E}$ is also separated into its components parallel and orthogonal to the azimuthal unit vector $\uvec{e}_\varphi$, so that 
\vspace{-3mm}
\begin{eqnarray}
\label{eq:E_comp_phi}
E_\varphi &=& \uvec{e}_\varphi \cdot \vec{E},\\
\label{eq:E_comp_rtheta}
|E_{r,\theta}| &=& |\uvec{e}_\varphi \times \vec{E}|. 
\end{eqnarray}
Due to symmetry, $E_\varphi=0$. On large distances from the antenna, 
the radial component vanishes, so that 
\begin{eqnarray}
\label{eq:measure_E}
\vec{E} \;\approx\;  {E}_\theta\, \uvec{e}_\theta.
\end{eqnarray}
%
We denote the complex electric far-field component from the references solution  $E^\text{ref}_\theta(\varphi,\theta)$, and when using an equivalent source $E^\text{}_\theta(\varphi,\theta)$.  
%
From the electric far-fields $E^\text{ref}_\theta(\varphi,\theta)$ and $E^\text{}_\theta(\varphi,\theta)$, we can evaluate both the magnitude error and the phase error. 
However, in this paper we only evaluate the magnitude error of the installed far-field, which is the most commonly used quantity to classify installed antennas. We define the relative magnitude error $\delta_\text{rel} |E_\theta(\varphi,\theta)|$ 
as
\begin{eqnarray}
\label{eq:measure_dE_rel}
\delta_\text{rel} |E_\theta(\varphi,\theta)| &=&  
\frac{|E_\theta(\varphi, \theta)| - |E_\theta^{\text{ref}}(\varphi,  \theta)|} {|E_\theta^{\text{ref}}(\varphi, \theta)|}
.
%
\end{eqnarray}

When calculating the current ${J}_t$, by using \eqref{eq:measure_Jt}, we evaluate ${H}_{\varphi}$ a small distance $d>0$ above the surface. The reason for this is numerical stability.  The field $\vec{H}$ will be zero inside the PEC structure, and by setting $d>0$ we assure that the evaluation points are not inside the structure. We use $d=0.5$~mm. 
%
By evaluating the far-field component ${E}_\theta$ we will be able to see systematic errors, which would be hidden in the normalization if e.g. the directivity was evaluated. 

Note that it is the field components $H_\varphi$ and $E_\theta$ in their respective regions that are used as inputs to the benchmarks in \eqref{eq:measure_dJt_rel}--\eqref{eq:measure_dJt_phase} and \eqref{eq:measure_dE_rel}. Hence, the approximations in \eqref{eq:measure_Jt} and \eqref{eq:measure_E} do not affect the accuracy of the benchmark\footnote{{For a rotational symmetric problem, as in this case, the expressions \eqref{eq:measure_Jt} and \eqref{eq:measure_E} are exact.}}.
The introduced notation, $\delta_\text{rel} |J_t(l)|$, $\delta \angle J_t(l)$ and $\delta_\text{rel} |E_\theta(\varphi,\theta)|$, is motivated by the physical relevance of these quantities.  

\subsection{Solvers and Simulation Settings}
\label{sec:MethodSolvers}
To evaluate the accuracy of the equivalent sources we need a reference case. We use a full wave solution {of} the detailed {antenna} model {in  Fig.~\ref{fig:main_structure}} as the reference. 
It is thus crucial that the reference solution is reliable. We here utilize two observations: that commercial solvers have high accuracy \cite{Vandenbosch2016} and that different numerical methods yield similar solutions. 

The equivalent sources are generated using FIT. 
In the evaluation of the equivalent sources, we examine their robustness, {with respect to design parameters}, by using them as excitations and solve in the full domain using FIT or MoM, as well as the asymptotic method Shooting-Bouncing-Rays (SBR).

All simulations are performed at  $10$~GHz 
using CST Microwave Studio (MWS) \cite{CSTMWS2016}. 
The frequency is chosen so that fields in the domain $D_0$ can be calculated with full-wave methods on a desktop computer. 
The size of the domains $D_0$ and $D_1$ is set so that they contain the structure of interest and $1.5 \, \lambda$ of space on each side. 


\vspace{0pt}
%

\section{Results}
\label{sec:Results}

\subsection{The Reference Solution}
\label{sec:Results0}

To examine the variability of the reference results, we use three numerical methods, FIT, MoM and FEM, applied to the model in Fig~\ref{fig:main_structure} with a model of the physical antenna. 

The magnitude of the tangential electrical current, $|J_t^\text{}(\ell)|$, along the line in Fig.~\ref{fig:main_structure_with_curve}(b), is depicted in Fig.~\ref{fig:res_hfield_phi_line_FIT_reference}.  
The currents calculated with different methods agree well, with a relative RMS difference between FIT and MoM currents of $3.4\,$\%. The difference between FIT and FEM is $9.3\,$\% and between MoM and FEM it is $9.9\,$\%. 

\begin{figure}[!t] 
	\centering
		\setlength{\unitlength}{1mm}
		\begin{picture}(80, 45)(0, -1.0)
			\put(0,0){
				\includegraphics[width=78mm, trim=34mm 108mm 40mm 114mm, clip] {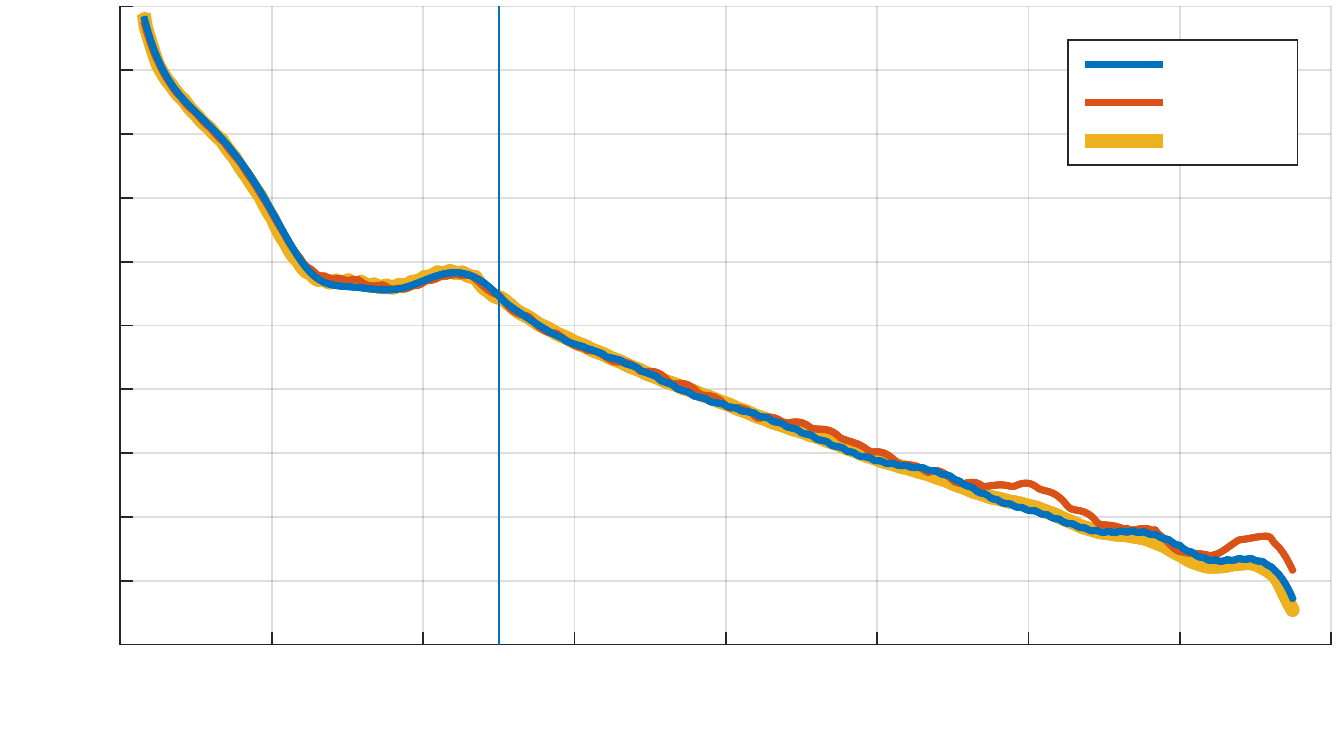}
			}
			\small 
		    \put(30,-1.5){\makebox(0,0)[lb]{\smash{Curve distance $\ell$ (mm)}}}%
		    \put( -0,15){\rotatebox{90}{\makebox(0,0)[lb]{\smash{$|J^\text{ref}_t(\ell)|$ (dBA/m)}}}}%

       		\put(7.5,2){\makebox(0,0)[lb]{\smash{0}}}%
    		\put(15, 2){\makebox(0,0)[lb]{\smash{20}}}%
    		\put(24, 2){\makebox(0,0)[lb]{\smash{40}}}%
    		\put(33, 2){\makebox(0,0)[lb]{\smash{60}}}%
    		\put(42, 2){\makebox(0,0)[lb]{\smash{80}}}%
    		\put(50, 2){\makebox(0,0)[lb]{\smash{100}}}%
    		\put(59, 2){\makebox(0,0)[lb]{\smash{120}}}%
    		\put(67, 2){\makebox(0,0)[lb]{\smash{140}}}%
    		\put(76, 2){\makebox(0,0)[lb]{\smash{160}}}%

    		\put( 7.0, 4.5){\makebox(0,0)[rb]{\smash{-35}}}%
    		\put( 7.0, 8.0){\makebox(0,0)[rb]{\smash{-30}}}%
    		\put( 7.0,12.0){\makebox(0,0)[rb]{\smash{-25}}}%
    		\put( 7.0,15.5){\makebox(0,0)[rb]{\smash{-20}}}%
    		\put( 7.0,19.5){\makebox(0,0)[rb]{\smash{-15}}}%
    		\put( 7.0,23.0){\makebox(0,0)[rb]{\smash{-10}}}%
    		\put( 7.0,26.5){\makebox(0,0)[rb]{\smash{-5}}}%
    		\put( 7.0,30.5){\makebox(0,0)[rb]{\smash{0}}}%
    		\put( 7.0,34.5){\makebox(0,0)[rb]{\smash{5}}}%
    		\put( 7.0,38.0){\makebox(0,0)[rb]{\smash{10}}}%
    		\put( 7.0,42.0){\makebox(0,0)[rb]{\smash{15}}}%
    		
			\footnotesize
		    \put(70.0,38.4){\makebox(0,0)[lb]{\smash{FIT}}}%
			\put(70.0,36.1){\makebox(0,0)[lb]{\smash{FEM}}}%
		    \put(70.0,33.8){\makebox(0,0)[lb]{\smash{MoM}}}%
   		\end{picture}			
	\caption{Reference case: The current $|J^\text{ref}_t(\ell)|$ 
	tangentially along the curve in Fig.~\ref{fig:main_structure_with_curve}(b), {solved with three different numerical methods}. 
	The vertical line at $l=50$~mm 
	marks the edge of the flat top.
		{
		}
	}
	\label{fig:res_hfield_phi_line_FIT_reference}
\end{figure}
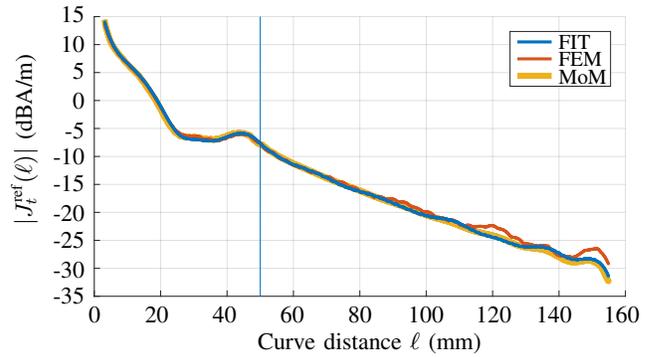
\begin{figure}[t] 
	\centering
		\setlength{\unitlength}{1mm}
		\begin{picture}(80, 45)(0, -1.0)
			\put(0,0){
					\includegraphics[width=78mm, trim=34mm 108mm 40mm 114mm, clip] {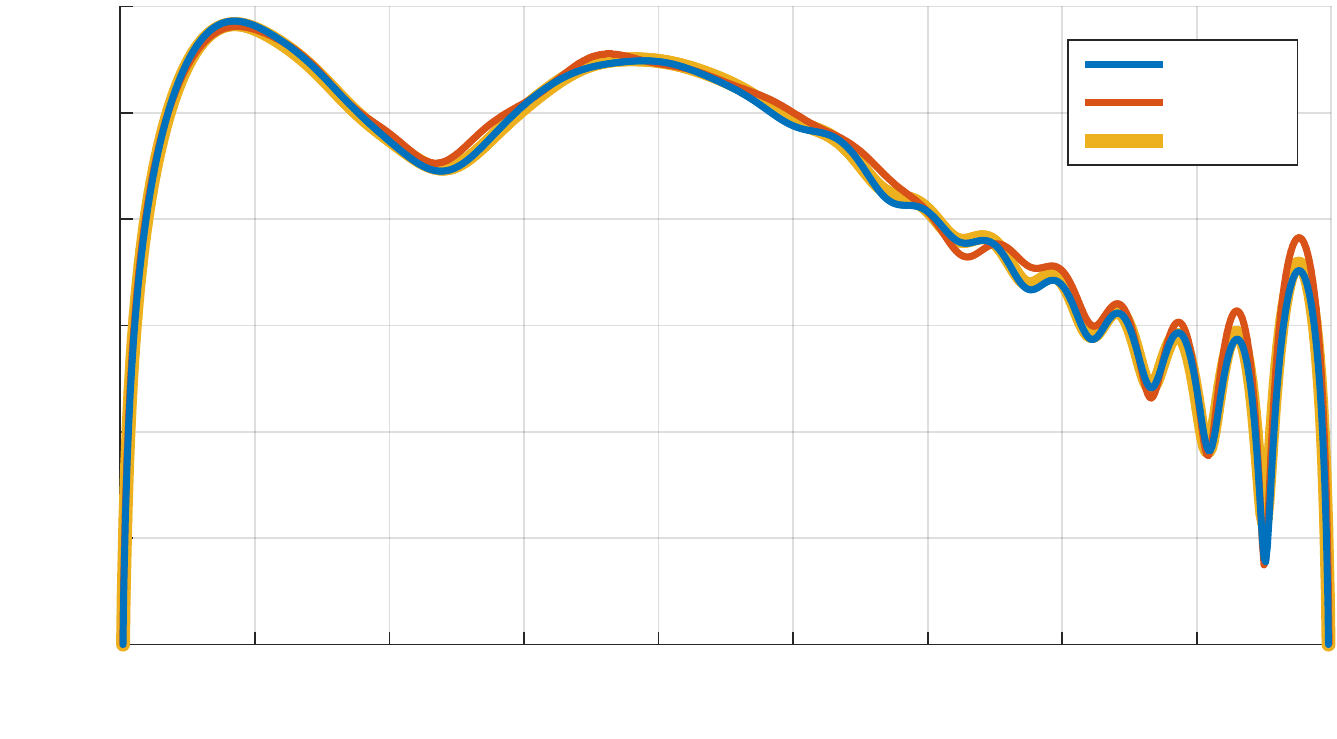}
			}
			\small 
		    \put(25,-1.5){\makebox(0,0)[lb]{\smash{Inclination angle  $\theta$ (degrees)}}}%
		    \put( -0,10){\rotatebox{90}{\makebox(0,0)[lb]{\smash{$|E^\text{ref}_\theta(90^\circ, \theta)|$ (dBV/m)}}}}%

       		\put(7.5,2){\makebox(0,0)[lb]{\smash{0}}}%
    		\put(14, 2){\makebox(0,0)[lb]{\smash{20}}}%
    		\put(22, 2){\makebox(0,0)[lb]{\smash{40}}}%
    		\put(30, 2){\makebox(0,0)[lb]{\smash{60}}}%
    		\put(38, 2){\makebox(0,0)[lb]{\smash{80}}}%
    		\put(45, 2){\makebox(0,0)[lb]{\smash{100}}}%
    		\put(53, 2){\makebox(0,0)[lb]{\smash{120}}}%
    		\put(61, 2){\makebox(0,0)[lb]{\smash{140}}}%
    		\put(68, 2){\makebox(0,0)[lb]{\smash{160}}}%
    		\put(76, 2){\makebox(0,0)[lb]{\smash{180}}}%

    		\put( 7.0,41.5){\makebox(0,0)[rb]{\smash{20}}}%
    		\put( 7.0,35.5){\makebox(0,0)[rb]{\smash{15}}}%
    		\put( 7.0,29.0){\makebox(0,0)[rb]{\smash{10}}}%
    		\put( 7.0,23){\makebox(0,0)[rb]{\smash{5}}}%
    		\put( 7.0,16.5){\makebox(0,0)[rb]{\smash{0}}}%
    		\put( 7.0,10.5){\makebox(0,0)[rb]{\smash{-5}}}%
    		\put( 7.0, 4.5){\makebox(0,0)[rb]{\smash{-10}}}%
    		
			\footnotesize
		    \put(70.0,38.4){\makebox(0,0)[lb]{\smash{FIT}}}%
			\put(70.0,36.1){\makebox(0,0)[lb]{\smash{FEM}}}%
		    \put(70.0,33.8){\makebox(0,0)[lb]{\smash{MoM}}}%
   		\end{picture}		
	\caption{Reference case: Installed electric far-field magnitude $|E_\theta^\text{ref}(90^\circ, \theta)|$, {solved with three different numerical methods}.
	}
	\label{fig:res_farfield_e-field_ref}
\end{figure}
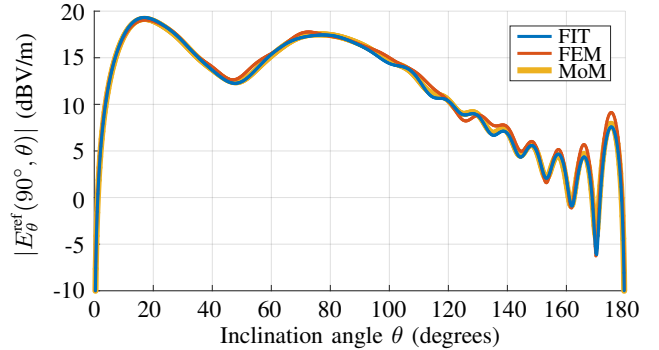

The installed far-field magnitude $|E_\theta^\text{}(\varphi, \theta)|$ is depicted in Fig.~\ref{fig:res_farfield_e-field_ref} for the azimuthal direction $\varphi=90^\circ$ and $\theta \in [0, 180^\circ]$. Again, the results from the numerical methods agree well, with a relative RMS difference between the far-fields calculated with FIT and MoM of $3.1\,$\%. The difference between FIT and FEM is $7.4\,$\% and between MoM and FEM it is $6.2\,$\%.

The agreement of the results from the different numerical methods {indicate} that the reference solutions are accurate. {We see in Fig.~	\ref{fig:res_hfield_phi_line_FIT_reference} that the solutions using MoM and FIT  conform best with the expected exponential decay of the current on the curved surface. Of these two solutions, with good mutual agreement, the FIT solution is chosen as the reference, since FIT is the native solver in CST MWS.}

We can relate some of the properties seen in 	Fig.~\ref{fig:res_hfield_phi_line_FIT_reference}--\ref{fig:res_farfield_e-field_ref} to the platform. The slope of surface current in Fig.~\ref{fig:res_hfield_phi_line_FIT_reference} for $\ell>50$~mm depends on the curvature of the sphere. The double beam at $\theta\approx 17^\circ$, $\theta\approx 78^\circ$, and the local minimum between, of the installed far-field in Fig.~\ref{fig:res_farfield_e-field_ref} is an effect of the flat top surface of the platform. A smaller radius of the flat top would lift the beams, i.e. decrease the inclination angle $\theta$. The oscillations for $ \theta \in (100^\circ, 180^\circ)$ in Fig.~\ref{fig:res_farfield_e-field_ref} {are} caused {by} reflections in the bottom edge of the platform $G$.

%

\subsection{Results for the Near-field Source Configuration}
\label{sec:Results1}

\begin{figure}[!t] 
	\centering
		\setlength{\unitlength}{1mm}
		\begin{picture}(80, 41)(0, 0)

			\put(0,24){
					\includegraphics[width=76mm, trim=46mm 163mm 46mm 106mm, clip] {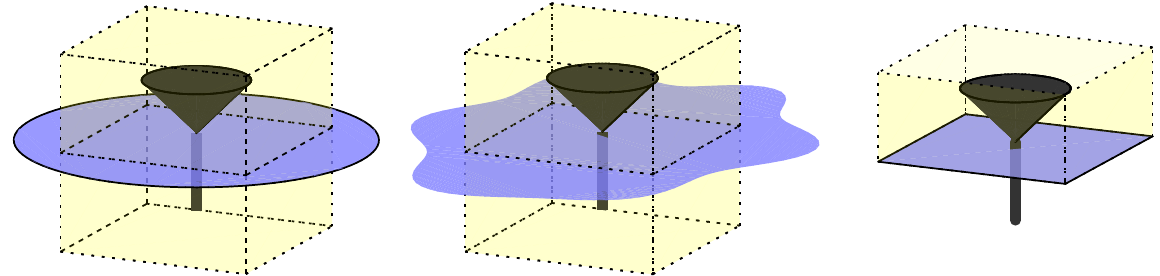}
			}
			\put(0,1){
					\includegraphics[width=76mm, trim=46mm 108mm 46mm 158mm, clip] {Fig_9_2016-09-15}
			}
			\small 
			\put(04,22){(a)}
			\put(31,22){(b)}			
			\put(58,22){(c)}
			\put(03,00){(d)}			
			\put(31,00){(e)}
			\put(58,00){(f)}	
					
		\end{picture}
	\caption{The configurations of ground plane (blue) and surface $\Gamma_e$ (yellow) when generating near-field sources; the ground-plane is (a) a thin $100$~mm diameter PEC plate, (b) a thin infinite PEC plate, (c) and (f) a thin $50\times 50$~mm$^2$ PEC plate, (d) a solid $100$~mm diameter, $25$~mm thick PEC plate, (e) the part of the  structure $G$ contained inside the interface $\Gamma_e$. 
	Note that the bottom of the surface $\Gamma_e$ coincide with the ground plane in (c), while it is $1$~mm above the ground plane i (f).
	}
	\label{fig:conf_eq_surf}
\end{figure}

The realization of near-field sources depend on the choice of the sub-domain $D_1$ and the placement of the surface $\Gamma_e$, see Fig.~\ref{fig:domains}, but also on the choice of ground-plane geometry. The aim here is to evaluate the freedom of choice with respect to 
these {geometrical parameters}. 
We consider six different configurations of surfaces $\Gamma_e$ and ground planes, as defined in Fig.~\ref{fig:conf_eq_surf}, with dimensions given in Table~\ref{tab:conf_eq_surf_z}. The configurations evaluated are  motivated briefly below.

%

\begin{table}[!t] 
  \centering
  \caption{Dimensions of surfaces $\Gamma_e$ for evaluated configurations.\vspace{-1mm}}
	\begin{tabular}{|l|ccc|}
		\hline
		Conf. 
		& $x$ & $y$ & $z$ \\
		\hline
		NFS (a) & $|x| = 25$~mm & $|y| = 25$~mm & $| z | = 25$~mm\\
		NFS (b) & $|x| = 25$~mm & $|y| = 25$~mm & $| z | = 25$~mm\\
		NFS (c) & $|x| = 25$~mm & $|y| = 25$~mm & $z=0$, $z= 25$~mm\\
		NFS (d) & $|x| = 25$~mm & $|y| = 25$~mm & $| z | = 25$~mm\\
		NFS (e) & $|x| = 25$~mm & $|y| = 25$~mm & $| z | = 25$~mm\\
		NFS (f) & $|x| = 25$~mm & $|y| = 25$~mm & $z=1$~\text{mm}, $z=25$~mm\\
		\hline
	\end{tabular}
	\label{tab:conf_eq_surf_z}
\end{table}

\begin{figure}[!t] 
	\centering
		\input{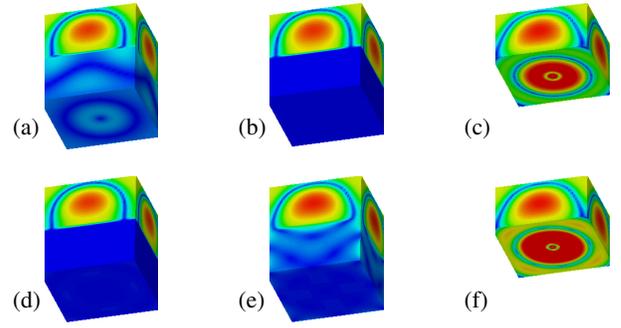}
	\caption{
	Tangential magnetic field magnitudes $|\uvec{n}\times\vec{H}|$ on
	$\Gamma_e$ for the configurations in Fig.~\ref{fig:conf_eq_surf}. 
	The logarithmic color scale ranges from $0$ to $1$~A/m.
	}%
	\label{fig:res_eq_surf_Htan_bottom}
\end{figure}

Configuration (a) is generated with a thin sheet PEC plate and has non-zero  currents also on the lower half of $\Gamma_e$, as can be seen in Fig.~\ref{fig:res_eq_surf_Htan_bottom}(a) and Fig.~\ref{fig:res_eq_surf_Etan_bottom}(a). The same applies to (e), because $\Gamma_e$ coincide with the PEC boundary. 
The infinite ground plane in (b) is essentially estimating the whole platform as a ground plane, whereas (a) and (d) account for the local geometry of the platform. The diameter of the circular PEC plate for the configurations (a) and (d) is $100$~mm ($3.3\,\lambda$) and corresponds to the flat top of the platform $G$, see Fig.~\ref{fig:main_structure}. 

Configuration (b) uses an infinite ground-plane. Hence, there will be no discontinuities in the surface currents on the ground plane. In configurations (d) and (e) the ground-plane is solid, resulting in a $90^\circ$ edge at radius $50$~mm on the ground plane. In the other configurations, (a), (c), and (f), the ground planes are thin sheets, resulting in sharp edges. 

One of the key features in (a) and (d) is that they capture a larger part of the platform geometry as compared to the other configurations. 
Configurations (c), (e), (f) all take a  ground plane with a side length $50$~mm ($1.7\, \lambda$) that correspond to the horizontal size of the equivalent surface $\Gamma_e$. The effect caused by the currents on the ground plane can be observed by comparing (c) and (f), since the surface $\Gamma_e$ coincide with the ground plane in (c) while it is $1$~mm above the ground plane in (f).
%
%
Note that the square ground plane in configurations (c), (e), (f) does not conform to the azimuthal symmetry of the original problem. 
However, the field solutions corresponding to configurations (c) and (e) show less asymmetry than (f), as can be seen in Fig.~\ref{fig:res_eq_surf_Etan_bottom}, possibly due to the effects of the ground plane that coincide with $\Gamma_e$ in (c) and (e), but not in (f). 
An advantage of setting the ground-plane size equal to the size of the surface $\Gamma_e$, as in (c), (e), and (f), is that the sub-domain $D_1$ is minimal, leading to shorter simulation times.

\begin{figure}[!t] 
	\centering
		\input{Etan_composite_NFS_bottom}
	\caption{
	Tangential electric field magnitudes $|\uvec{n}\times\vec{E}|$ on
	$\Gamma_e$ for the con\-figura\-tions in Fig.~\ref{fig:conf_eq_surf}. The logarithmic color scale ranges from $0$ to $320$~V/m.
	}%
	\label{fig:res_eq_surf_Etan_bottom}
\end{figure}

The work flow described in Section~\ref{sec:workflow_NFS} is followed when generating and using the near-field sources. Since the reference solution was solved with FIT, we use FIT again to solve the problem with the near-field sources imprinted in the platform model. 

The resulting near-field source representations of the antenna are depicted in Fig.~\ref{fig:res_eq_surf_Htan_bottom}--\ref{fig:res_eq_surf_Etan_bottom}, for each of the configurations used. The strong fields on the bottom surface of (c) and (f) is due to the coaxial feed cable that penetrates the surface $\Gamma_e$.

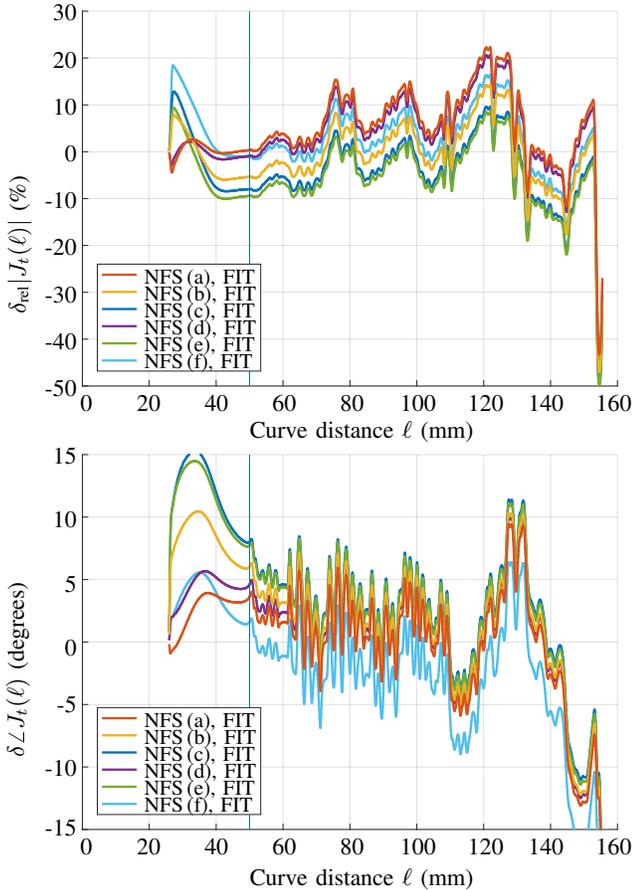
\begin{figure}[!t] 
	\centering
	\setlength{\unitlength}{1mm}
		\begin{picture}(80, 57)(0, 0)
			\put(0,2){
				\includegraphics[width=78mm, trim=35mm 98mm 40mm 102mm, clip] {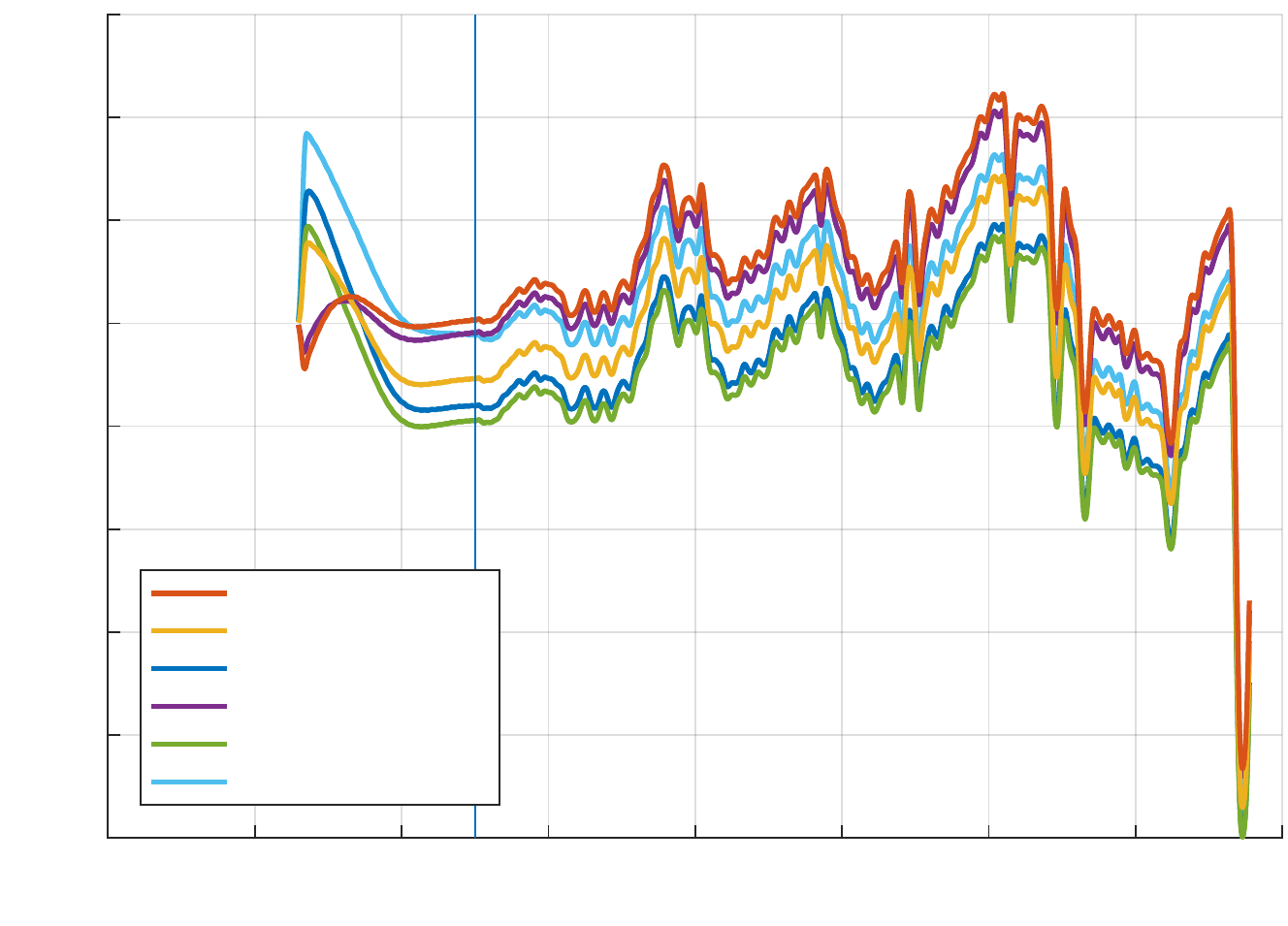}
			}
			\small 
		    \put(30,0.5){\makebox(0,0)[lb]{\smash{Curve distance $\ell$ (mm)}}}%
		    \put( 0.2,17){\rotatebox{90}{\makebox(0,0)[lb]{\smash{$\delta_\text{rel} |J_t(\ell)|$ (\%)}}}}%

       		\put(7.5,4){\makebox(0,0)[lb]{\smash{0}}}%
    		\put(15, 4){\makebox(0,0)[lb]{\smash{20}}}%
    		\put(24, 4){\makebox(0,0)[lb]{\smash{40}}}%
    		\put(33, 4){\makebox(0,0)[lb]{\smash{60}}}%
    		\put(42, 4){\makebox(0,0)[lb]{\smash{80}}}%
    		\put(50, 4){\makebox(0,0)[lb]{\smash{100}}}%
    		\put(59, 4){\makebox(0,0)[lb]{\smash{120}}}%
    		\put(67, 4){\makebox(0,0)[lb]{\smash{140}}}%
    		\put(76, 4){\makebox(0,0)[lb]{\smash{160}}}%
		
    		\put( 7.0,56.0){\makebox(0,0)[rb]{\smash{30}}}%
    		\put( 7.0,49.75){\makebox(0,0)[rb]{\smash{20}}}%
    		\put( 7.0,43.5){\makebox(0,0)[rb]{\smash{10}}}%
    		\put( 7.0,37.25){\makebox(0,0)[rb]{\smash{0}}}%
    		\put( 7.0,31.00){\makebox(0,0)[rb]{\smash{-10}}}%
    		\put( 7.0,24.75){\makebox(0,0)[rb]{\smash{-20}}}%
    		\put( 7.0,18.50){\makebox(0,0)[rb]{\smash{-30}}}%
    		\put( 7.0,12.25){\makebox(0,0)[rb]{\smash{-40}}}%
    		\put( 7.0, 6.00){\makebox(0,0)[rb]{\smash{-50}}}%
    		
    		\footnotesize
		    \put(16.0,21.00){\makebox(0,0)[lb]{\smash{NFS$\,$(a), FIT}}}%
			\put(16.0,18.75){\makebox(0,0)[lb]{\smash{NFS$\,$(b), FIT}}}%
			\put(16.0,16.50){\makebox(0,0)[lb]{\smash{NFS$\,$(c), FIT}}}%
		    \put(16.0,14.25){\makebox(0,0)[lb]{\smash{NFS$\,$(d), FIT}}}%
		    \put(16.0,12.00){\makebox(0,0)[lb]{\smash{NFS$\,$(e), FIT}}}%
		    \put(16.0, 9.75){\makebox(0,0)[lb]{\smash{NFS$\,$(f), FIT}}}%
   		\end{picture}			
\\
		\setlength{\unitlength}{1mm}
		\begin{picture}(80, 59)(0, -2)
			\put(0,0){
				\includegraphics[width=78mm, trim=35mm 98mm 40mm 102mm, clip] {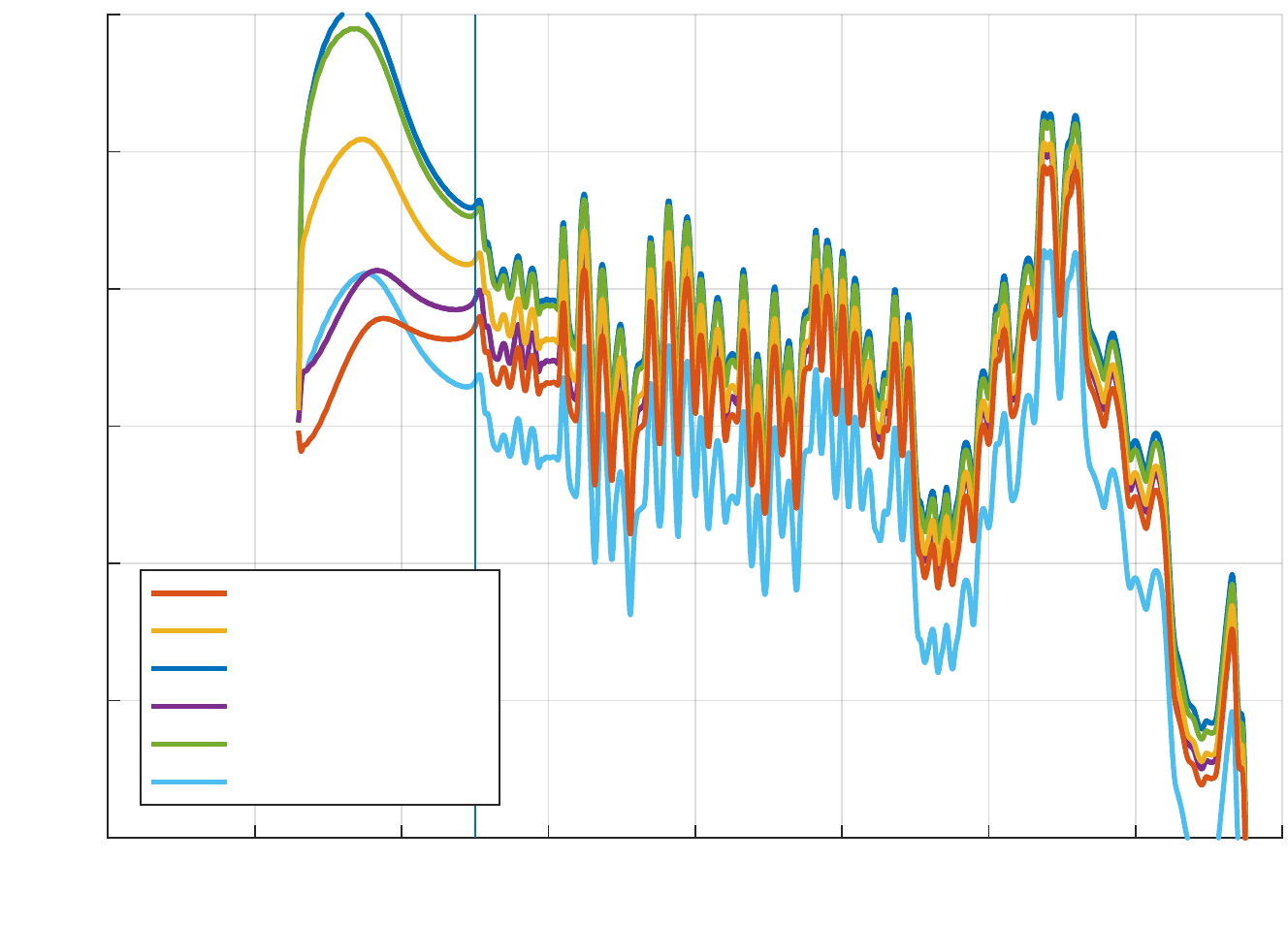}
			}
			\small 
		    \put(30,-2){\makebox(0,0)[lb]{\smash{Curve distance $\ell$ (mm)}}}%
		    \put( -0,15){\rotatebox{90}{\makebox(0,0)[lb]{\smash{$\delta\protect\angle J_t(\ell)$ (degrees)}}}}%

       		\put(7.5,2){\makebox(0,0)[lb]{\smash{0}}}%
    		\put(15, 2){\makebox(0,0)[lb]{\smash{20}}}%
    		\put(24, 2){\makebox(0,0)[lb]{\smash{40}}}%
    		\put(33, 2){\makebox(0,0)[lb]{\smash{60}}}%
    		\put(42, 2){\makebox(0,0)[lb]{\smash{80}}}%
    		\put(50, 2){\makebox(0,0)[lb]{\smash{100}}}%
    		\put(59, 2){\makebox(0,0)[lb]{\smash{120}}}%
    		\put(67, 2){\makebox(0,0)[lb]{\smash{140}}}%
    		\put(76, 2){\makebox(0,0)[lb]{\smash{160}}}%

    		\put( 7.0, 4.5){\makebox(0,0)[rb]{\smash{-15}}}%
    		\put( 7.0,12.5){\makebox(0,0)[rb]{\smash{-10}}}%
    		\put( 7.0,21.0){\makebox(0,0)[rb]{\smash{-5}}}%
    		\put( 7.0,28.5){\makebox(0,0)[rb]{\smash{0}}}%
    		\put( 7.0,37.0){\makebox(0,0)[rb]{\smash{5}}}%
    		\put( 7.0,45.5){\makebox(0,0)[rb]{\smash{10}}}%
    		\put( 7.0,54.0){\makebox(0,0)[rb]{\smash{15}}}%
    		
    		\footnotesize
		    \put(16.0,19.00){\makebox(0,0)[lb]{\smash{NFS$\,$(a), FIT}}}%
			\put(16.0,16.75){\makebox(0,0)[lb]{\smash{NFS$\,$(b), FIT}}}%
			\put(16.0,14.50){\makebox(0,0)[lb]{\smash{NFS$\,$(c), FIT}}}%
		    \put(16.0,12.25){\makebox(0,0)[lb]{\smash{NFS$\,$(d), FIT}}}%
		    \put(16.0,10.00){\makebox(0,0)[lb]{\smash{NFS$\,$(e), FIT}}}%
		    \put(16.0, 7.75){\makebox(0,0)[lb]{\smash{NFS$\,$(f), FIT}}}%
   		\end{picture}			
	\caption{
	The relative surface current  magnitude error $\delta_\text{rel} |J_{t}(\ell)|$
	(top) and  surface current phase error $\delta\protect\angle J_{t}(\ell)$  (bottom), for different near-field source configurations, see Fig.~\ref{fig:conf_eq_surf}. The errors are evaluated along the curve in Fig.~\ref{fig:main_structure_with_curve}(b). %
	The vertical line at $\ell=50$~mm marks the edge of the flat top of the geometry. 
	}
	\label{fig:res_currents_H2_eq_surf}
\end{figure}
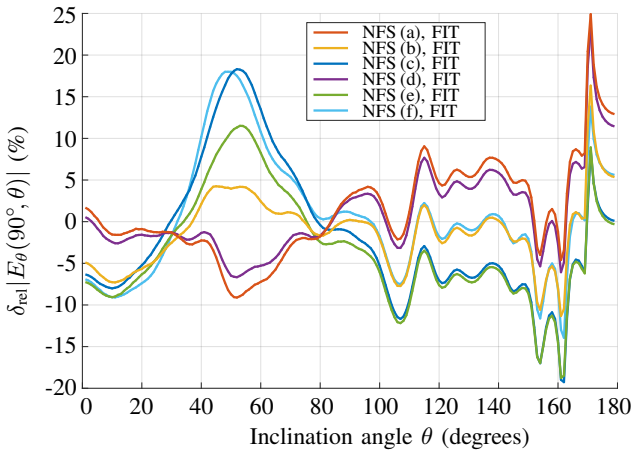
\begin{figure}[!t] 
	\centering
		\setlength{\unitlength}{1mm}
		\begin{picture}(80, 59)(0, -2)
			\put(0,0){
				\includegraphics[width=78mm, trim=35mm 98mm 40mm 102mm, clip] {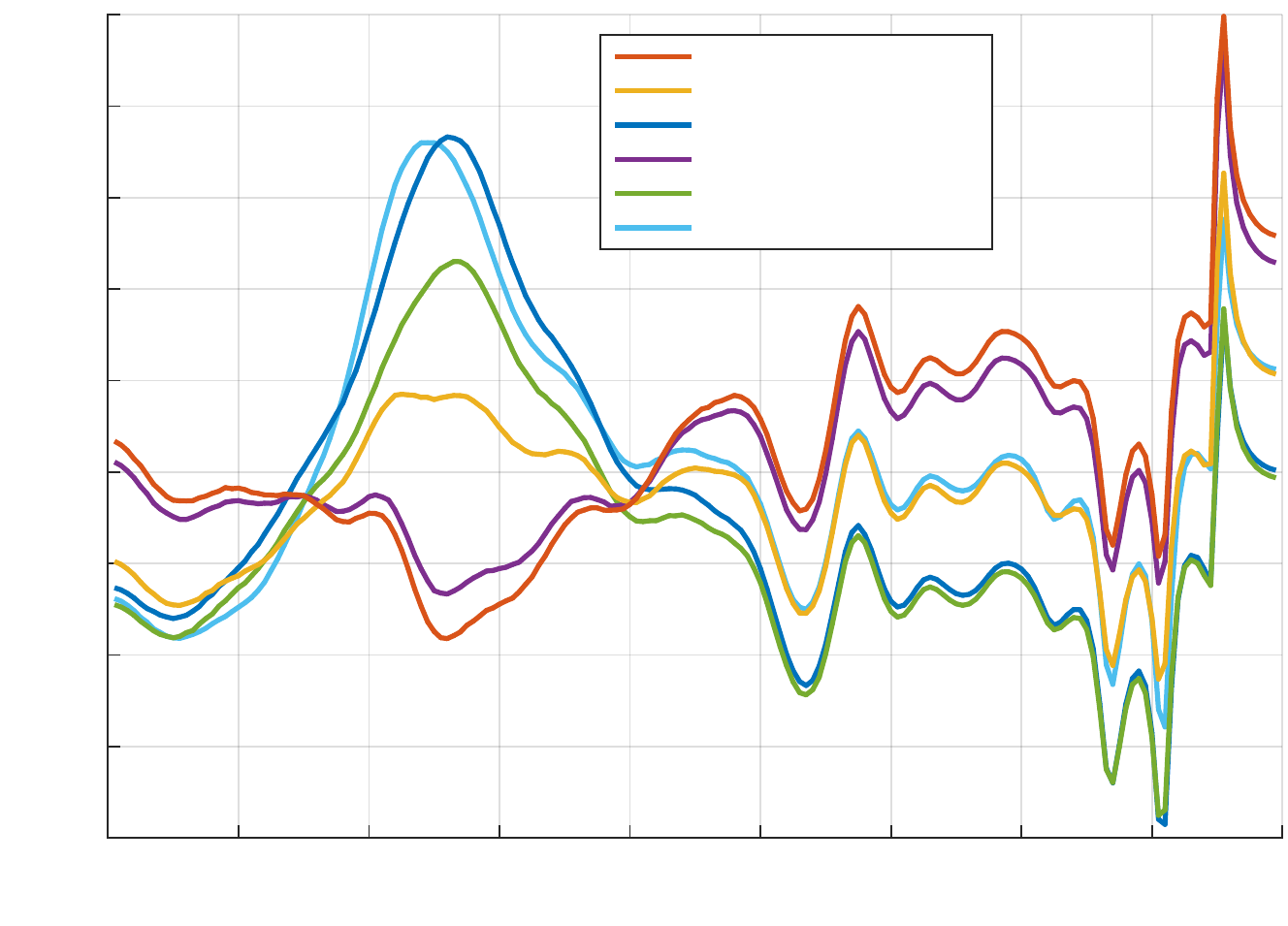}
			}
			
			\small 
		    \put(30,-2){\makebox(0,0)[lb]{\smash{Inclination angle $\theta$ (degrees)}}}%
		    \put( -0,15){\rotatebox{90}{\makebox(0,0)[lb]{\smash{$\delta_\text{rel} |E_\theta(90^\circ\!, \theta)|$ (\%)}}}}%

       		\put(7.5,2){\makebox(0,0)[lb]{\smash{0}}}%
    		\put(14, 2){\makebox(0,0)[lb]{\smash{20}}}%
    		\put(22, 2){\makebox(0,0)[lb]{\smash{40}}}%
    		\put(30, 2){\makebox(0,0)[lb]{\smash{60}}}%
    		\put(38, 2){\makebox(0,0)[lb]{\smash{80}}}%
    		\put(45, 2){\makebox(0,0)[lb]{\smash{100}}}%
    		\put(53, 2){\makebox(0,0)[lb]{\smash{120}}}%
    		\put(60, 2){\makebox(0,0)[lb]{\smash{140}}}%
    		\put(68, 2){\makebox(0,0)[lb]{\smash{160}}}%
    		\put(76, 2){\makebox(0,0)[lb]{\smash{180}}}%

    		\put( 7.0,54.0){\makebox(0,0)[rb]{\smash{25}}}%
    		\put( 7.0,48.5){\makebox(0,0)[rb]{\smash{20}}}%
    		\put( 7.0,43.0){\makebox(0,0)[rb]{\smash{15}}}%
    		\put( 7.0,37.5){\makebox(0,0)[rb]{\smash{10}}}%
    		\put( 7.0,32.0){\makebox(0,0)[rb]{\smash{5}}}%
    		\put( 7.0,26.5){\makebox(0,0)[rb]{\smash{0}}}%
    		\put( 7.0,21.0){\makebox(0,0)[rb]{\smash{-5}}}%
    		\put( 7.0,15.5){\makebox(0,0)[rb]{\smash{-10}}}%
    		\put( 7.0,10.0){\makebox(0,0)[rb]{\smash{-15}}}%
    		\put( 7.0, 4.5){\makebox(0,0)[rb]{\smash{-20}}}%

    		\scriptsize
		    \put(45,52.0){\makebox(0,0)[lb]{\smash{NFS$\,$(a), FIT}}}%
			\put(45,49.8){\makebox(0,0)[lb]{\smash{NFS$\,$(b), FIT}}}%
			\put(45,47.8){\makebox(0,0)[lb]{\smash{NFS$\,$(c), FIT}}}%
		    \put(45,45.7){\makebox(0,0)[lb]{\smash{NFS$\,$(d), FIT}}}%
		    \put(45,43.6){\makebox(0,0)[lb]{\smash{NFS$\,$(e), FIT}}}%
			\put(45,41.5){\makebox(0,0)[lb]{\smash{NFS$\,$(f), FIT}}}%
   		\end{picture}		
	\caption{
	Relative installed electric far-field errors $\delta_\text{rel} |E_\theta(90^\circ,\theta)|$  using the equivalent near-field sources illustrated in Fig.~\ref{fig:conf_eq_surf}.
	}	
	\label{fig:res_farfields_eq_surf_full_view}
\end{figure}

The accuracy is evaluated by the tangential current errors $\delta_\text{rel} |J_t(\ell)|$ and $\delta_\text{} \angle J_t(\ell)$, according to \eqref{eq:measure_dJt_rel} and \eqref{eq:measure_dJt_phase}. They are presented in Table~\ref{tab:rms_eq_surf} as RMS errors and depicted in Fig.~\ref{fig:res_currents_H2_eq_surf} for the interval $\ell\in(25, 155]$~mm, i.e. outside the equivalent surface $\Gamma_e$. 
%
The installed far-field errors $\delta_\text{rel} |E_\theta(90^\circ\!, \theta)|$, according to \eqref{eq:measure_dE_rel}, are depicted in Fig.~\ref{fig:res_farfields_eq_surf_full_view} and the RMS errors for $\theta\in(0, 180^\circ)$ are listed in Table~\ref{tab:rms_eq_surf}.

\begin{table}[!t] 
  \centering
	  \caption{Root-mean-square errors using near-field source and FIT.\vspace{-1mm}} 
	\begin{tabular}{|l||r|r|r|}
		\hline
		\multirow{2}{*}{
				 	\hspace{-2mm}
				 	\parbox[lb]{8mm}{Configu\-ration}
				 }
			  & \multicolumn{3}{c|}{RMS (linear scale)}\\
  	\cline{2-4}
		& \parbox[t]{21mm}{$\delta_\text{rel} |J_{t}(\ell)|$,\\
						   $\ell\in(25, 155]$ mm} 
		& \parbox[t]{21mm}{$\delta\angle J_{t}(\ell)$, \\
						   $\ell\in(25, 155]$ mm} 
		& \parbox[t]{18mm}{$\delta_\text{rel} |E_{\theta}(90^\circ\!,\theta)|$,\\
						   $\theta\in(0, 180^\circ)$}  
		\\
		\hline
		%
		%
		    
		NFS (a) &  $9.2\,$\%  
				& $4.0^\circ$ 
				& $6.0\,$\%   
				\\
		NFS (b) & $12\,$\%    
				& $7.1^\circ$ 
				& $4.4\,$\%   
				\\
		NFS (c) & $18\,$\%    
				& $7.2^\circ$ 
				& $8.4\,$\%   
				\\
		NFS (d) &  $8.0\,$\%  
				& $4.5^\circ$ 
				& $5.1\,$\%   
				\\
		NFS (e) & $19\,$\%    
				& $6.9^\circ$ 
				& $7.4\,$\%   
				\\
		NFS (f) & $17\,$\%    
				& $3.8^\circ$ 
				& $7.3\,$\%   
				\\
		\hline
		\end{tabular}
	\label{tab:rms_eq_surf}
\end{table}

The best behavior in terms of $\delta_\text{rel} |J_{t}(\ell)|$ is given by configurations (d) and (a) with $8.0\,$\% and $9.2\,$\% relative error, respectively, whereas (c), (e), (f) are the worst, with about twice as large relative errors as (a) and (d). The correct local geometry of the ground-planes in (a) and (d) is important for the accuracy of the surface currents.

For the installed far-field in Fig.~\ref{fig:res_farfields_eq_surf_full_view}, we see that the variation is smaller between the different configurations, as compared to the current error in Fig.~\ref{fig:res_currents_H2_eq_surf}. This is expected {due to the smoothing effect for far-fields}.
The smallest RMS error $\delta_\text{rel} |E_\theta(90^\circ,\theta)|$ are for (b) with $4.4\,$\% and (d) with $5.1\,$\%{, which are comparable with the variations of the reference solutions on $3.1\,$\%}. 
In Fig.~\ref{fig:res_farfields_eq_surf_full_view} the difference between configuration is particularly large in the region $\theta\in [30^\circ,  80^\circ]$, motivating a closer study. The max-norm deviations in this region range from $4^\circ$ with (b) to $18^\circ$ with (c) and (f). We see that the {edge of the ground-plane} seems to play an important role for the accuracy in this region, where (b) performs best (no discontinuity), and the configuration with a square thin sheet ground-plane, (c) and (f), give the least accurate results. Comparing (a) and (d), we see that (d) with a solid ground-plane ($90^\circ$ edge) is {more accurate} than (a) with a thin ground-plane (sharp edge). We see the same pattern when comparing (c) with (e); the solid ground-plane performs better than the thin sheet ground-plane with a sharp edge. {The size of the ground plane also plays a role. A small ground plane gives a lifting effect of the pattern from the horizontal plane (as discussed in Sec.~\ref{sec:Results0}). The large errors for (c), (e), (f) is partly caused by this effect, where the beam maximum is shifted from $\theta \approx 78^\circ$ to $\theta \approx 72^\circ$.}
The computationally simple configuration (b) with an infinite ground-plane gives accurate results, especially for the installed far-fields.

The expected higher sensitivity of the currents as compared with the far-field is clearly observed with a difference of a factor of about two for RMS errors. 
Compared with the estimated RMS uncertainty in the reference solutions in Section~\ref{sec:Results0} ($3.4\,$\% for the current magnitude and $3.1\,$\% for the installed far-field magnitudes), we see {in Table~\ref{tab:rms_eq_surf}} that the near-field sources {increase the current magnitude uncertainty with a factor of $2.3$--$5.6$ 
and the far-field magnitude uncertainty with a factor of $1.4$--$2.7$ 
.} 


To conclude this section, we note that the RMS errors of the currents, is about $9\,$\% {for the best case} (see Table~\ref{tab:rms_eq_surf}), with max-norm deviations up to $\pm 20\,$\% for $\ell\in (25,150]$~mm, rising up to $\pm 50\,$\% close to the bottom platform edge at $\ell = 155$~mm (see Fig.~\ref{fig:res_currents_H2_eq_surf}). Similarly, the phase has about $5^\circ$ RMS error for the best configurations (see Table~\ref{tab:rms_eq_surf}), with max-norm deviation up to $\pm 20^\circ$ (see Fig.~\ref{fig:res_currents_H2_eq_surf}). 
For the installed far-fields, we note that the RMS errors are about $5\,$\% for the {best} cases, with corresponding max-norm deviations up to $\pm 10\,$\% over $\theta\in [0,160^\circ]$. We note that the RMS errors {of the far-field} vary with a factor of two between the most and least accurate configurations.

\subsection{Results for Far-field Source Configuration}
\label{sec:Results2}
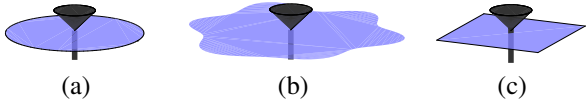
\begin{figure}[!t] 
	\centering
		\setlength{\unitlength}{1mm}
			\begin{picture}(80, 13)(0, 0)
				\put(0,3){
						\includegraphics[width=78mm, trim=52mm 170mm 46mm 113mm, clip] {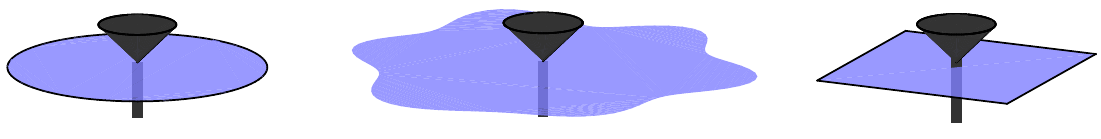}
				}
				\put(09.2,00){(a)}
				\put(38,00){(b)}			
				\put(67.2,00){(c)}			
		\end{picture}
	\caption{The configurations of ground plane when generating far-field sources; (a) a thin 100~mm diameter PEC plate, (b) a thin infinite PEC plate, and (c) a thin $50 \times 50$~mm$^2$ PEC plate.
	}%
	\label{fig:conf_eq_farfield}
\end{figure}

\begin{figure}[!t] 
	\centering
	\setlength{\unitlength}{1mm}
		\begin{picture}(80, 42)(0, 0)
			\put(2.5,2.5){
				\includegraphics[width=78mm, trim=28mm 112mm 20mm 102mm, clip] {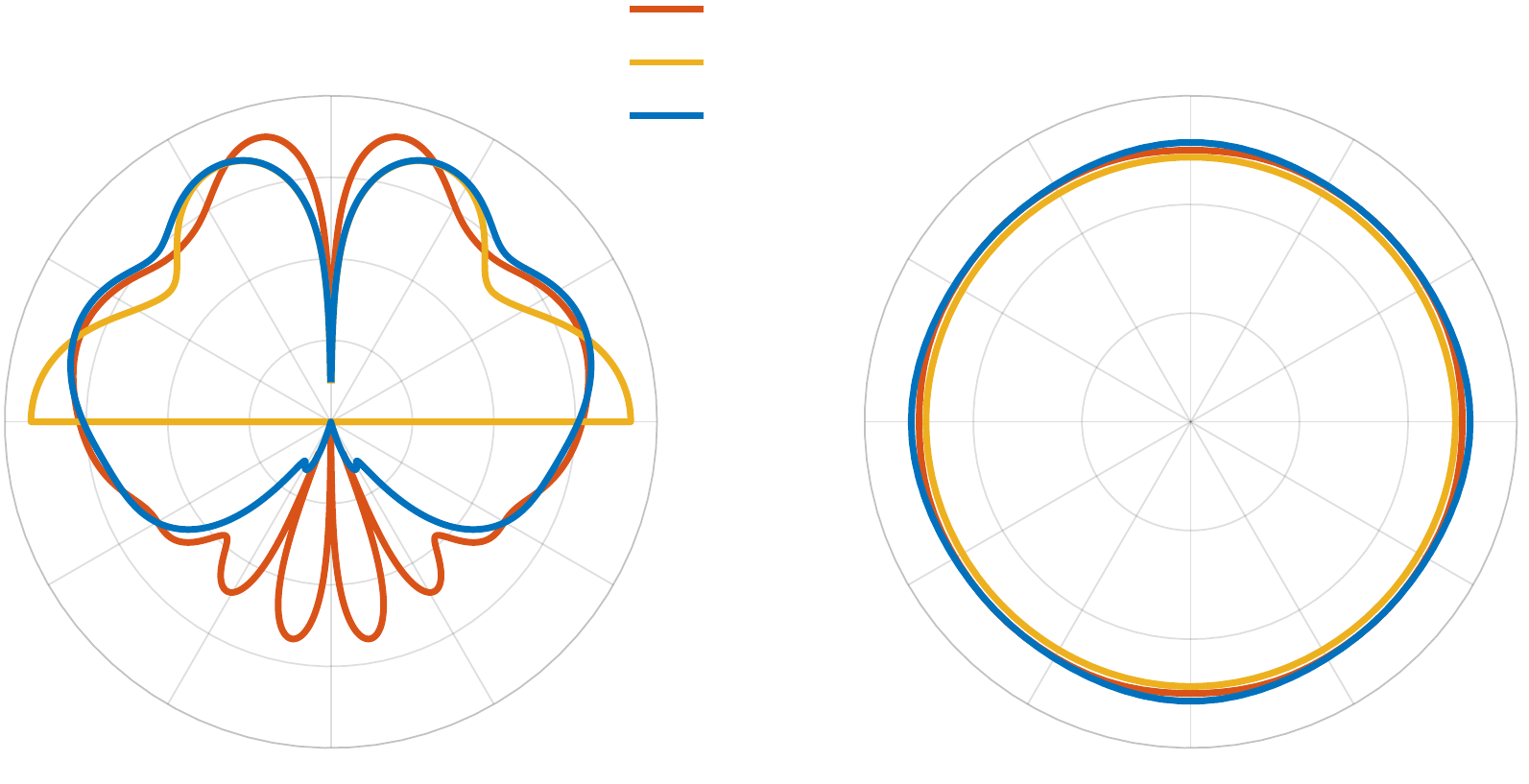}		
		   	}
	\put(2,0){\color[rgb]{0,0,0}\makebox(0,0)[lb]{\smash{(a)}}}%
	\put(45,0){\color[rgb]{0,0,0}\makebox(0,0)[lb]{\smash{(b)}}}%
		   	
	\footnotesize
	\put(40.5,41.30){\color[rgb]{0,0,0}\makebox(0,0)[lb]{\smash{FFS$\,$(a)}}}%
	\put(40.5,38.45){\color[rgb]{0,0,0}\makebox(0,0)[lb]{\smash{FFS$\,$(b)}}}%
	\put(40.5,35.60){\color[rgb]{0,0,0}\makebox(0,0)[lb]{\smash{FFS$\,$(c)}}}%
	\linethickness{0.10mm}
	\put(34,44){\color[rgb]{0.3,0.3,0.3}\line(1,0){16}}
	\put(34,35){\color[rgb]{0.3,0.3,0.3}\line(1,0){16}}
	\put(34,44){\color[rgb]{0.3,0.3,0.3}\line(0,-1){9}}
	\put(50,44){\color[rgb]{0.3,0.3,0.3}\line(0,-1){9}}
	
 	\footnotesize    	
    \put(14,22){\color[rgb]{0.1,0.1,0.1}\rotatebox{0}{\makebox(0,0)[lb]{\smash{$-20$}}}}%
    \put(10,24){\color[rgb]{0.1,0.1,0.1}\rotatebox{0}{\makebox(0,0)[lb]{\smash{$-10$}}}}%
    \put(08,25){\color[rgb]{0.1,0.1,0.1}\rotatebox{0}{\makebox(0,0)[lb]{\smash{$0$}}}}%
    \put(03,26){\color[rgb]{0.1,0.1,0.1}\rotatebox{0}{\makebox(0,0)[lb]{\smash{$10$}}}}%

    \put(18,38.5){\color[rgb]{0.1,0.1,0.1}\rotatebox{0}{\makebox(0,0)[lb]{\smash{$\theta=0$}}}}%
    \put(9.5,36){\color[rgb]{0.1,0.1,0.1}\rotatebox{0}{\makebox(0,0)[lb]{\smash{$30$}}}}%
    \put(28.5,36){\color[rgb]{0.1,0.1,0.1}\rotatebox{0}{\makebox(0,0)[lb]{\smash{$30$}}}}%
    
    \put(03,29.0){\color[rgb]{0.1,0.1,0.1}\rotatebox{0}{\makebox(0,0)[lb]{\smash{$60$}}}}%
    \put(35.5,29.0){\color[rgb]{0.1,0.1,0.1}\rotatebox{0}{\makebox(0,0)[lb]{\smash{$60$}}}}%
    
    \put(0,20){\color[rgb]{0.1,0.1,0.1}\rotatebox{0}{\makebox(0,0)[lb]{\smash{$90$}}}}%
    \put(38,20){\color[rgb]{0.1,0.1,0.1}\rotatebox{0}{\makebox(0,0)[lb]{\smash{$90$}}}}%
    
    \put(1,10.5){\color[rgb]{0.1,0.1,0.1}\rotatebox{0}{\makebox(0,0)[lb]{\smash{$120$}}}}%
    \put(35,10.5){\color[rgb]{0.1,0.1,0.1}\rotatebox{0}{\makebox(0,0)[lb]{\smash{$120$}}}}%
    
    \put(9,3.5){\color[rgb]{0.1,0.1,0.1}\rotatebox{0}{\makebox(0,0)[lb]{\smash{$150$}}}}%
    \put(28,3.5){\color[rgb]{0.1,0.1,0.1}\rotatebox{0}{\makebox(0,0)[lb]{\smash{$150$}}}}%

    \put(18,1){\color[rgb]{0.1,0.1,0.1}\rotatebox{0}{\makebox(0,0)[lb]{\smash{$180$}}}}%

 	\footnotesize    	
    \put(57,22){\color[rgb]{0.1,0.1,0.1}\rotatebox{0}{\makebox(0,0)[lb]{\smash{$-5$}}}}%
    \put(54,23.5){\color[rgb]{0.1,0.1,0.1}\rotatebox{0}{\makebox(0,0)[lb]{\smash{$0$}}}}%
    \put(48,26){\color[rgb]{0.1,0.1,0.1}\rotatebox{0}{\makebox(0,0)[lb]{\smash{$5$}}}}%

    \put(61,38.5){\color[rgb]{0.1,0.1,0.1}\rotatebox{0}{\makebox(0,0)[lb]{\smash{$\varphi=0$}}}}%
    \put(73,36){\color[rgb]{0.1,0.1,0.1}\rotatebox{0}{\makebox(0,0)[lb]{\smash{$30$}}}}%
    \put(79.5,29){\color[rgb]{0.1,0.1,0.1}\rotatebox{0}{\makebox(0,0)[lb]{\smash{$60$}}}}%
    \put(82,20){\color[rgb]{0.1,0.1,0.1}\rotatebox{0}{\makebox(0,0)[lb]{\smash{$90$}}}}%
    \put(78.5,10){\color[rgb]{0.1,0.1,0.1}\rotatebox{0}{\makebox(0,0)[lb]{\smash{$120$}}}}%
    \put(71,3){\color[rgb]{0.1,0.1,0.1}\rotatebox{0}{\makebox(0,0)[lb]{\smash{$150$}}}}%
    \put(62,1){\color[rgb]{0.1,0.1,0.1}\rotatebox{0}{\makebox(0,0)[lb]{\smash{$180$}}}}%
    \put(51.5,3.5){\color[rgb]{0.1,0.1,0.1}\rotatebox{0}{\makebox(0,0)[lb]{\smash{$210$}}}}%
    \put(45,10.5){\color[rgb]{0.1,0.1,0.1}\rotatebox{0}{\makebox(0,0)[lb]{\smash{$240$}}}}%
    \put(43,20){\color[rgb]{0.1,0.1,0.1}\rotatebox{0}{\makebox(0,0)[lb]{\smash{$270$}}}}%
    \put(45,29.5){\color[rgb]{0.1,0.1,0.1}\rotatebox{0}{\makebox(0,0)[lb]{\smash{$300$}}}}%
    \put(52.5,36){\color[rgb]{0.1,0.1,0.1}\rotatebox{0}{\makebox(0,0)[lb]{\smash{$330$}}}}%
		\end{picture}	
	\caption{Realized gain of far-field sources for the configurations illustrated in Fig.~\ref{fig:conf_eq_farfield}, for (a) azimuths $\varphi=\pm 90^\circ$ and inclinations $\theta\in [\,0, 180^\circ]$, and (b) azimuths $\varphi=\in [\,0, 180^\circ]$ and inclination $\theta = 70^\circ$.
	}%
	\label{fig:res_eq_farfield}
\end{figure}
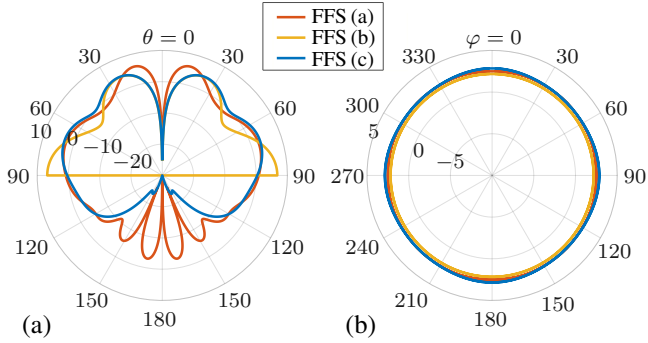

When generating the far-field sources, we consider three configurations, all defined in Fig.~\ref{fig:conf_eq_farfield}. 
From these configurations, we use FIT to calculate their far-field patterns. The results, which are used as far-field sources, are depicted in Fig.~\ref{fig:res_eq_farfield}(a) for the vertical cut $\varphi=\pm 90^\circ$ and Fig.~\ref{fig:res_eq_farfield}(b) for the horizontal cut $\theta=70^\circ$. The inclination $\theta=70^\circ$ is depicted since the three far-field patterns have similar magnitude {making it easy to compare the curves and identify asymmetries}. 
Note that configuration (b), with an infinite ground plane, results in a far-field pattern that is identical to zero in the lower hemisphere, $\theta > 90^\circ$. Configuration (c) will not preserve the symmetry in $\varphi$ of the original problem, but the effect is small, as can be seen in Fig.~\ref{fig:res_eq_farfield}(b).

The work flow described in Section~\ref{sec:workflow_FFS} is used for generating and using the far-field sources. 
Because of the rotational symmetry of the platform geometry $G$, the far-field source is placed on the symmetry axis $x=0$, $y=0$. {In contrast}, the position $h$ on the vertical axis is not trivial. Compared to e.g. a geometrical theory of diffraction (GTD) formulation \cite{PathakWang1981}, the source, in that case a dipole moment, can be placed both on the conducting surface or above it.
We investigate four cases of the design parameter $h=(0, 2, 4, 10)\,$mm, i.e. the distance above the flat platform top, see Fig.~\ref{fig:domains}. 
The resulting problem is solved with MoM, since FIT in CST Microwave Studio \cite{CSTMWS2016} cannot handle far-field sources.

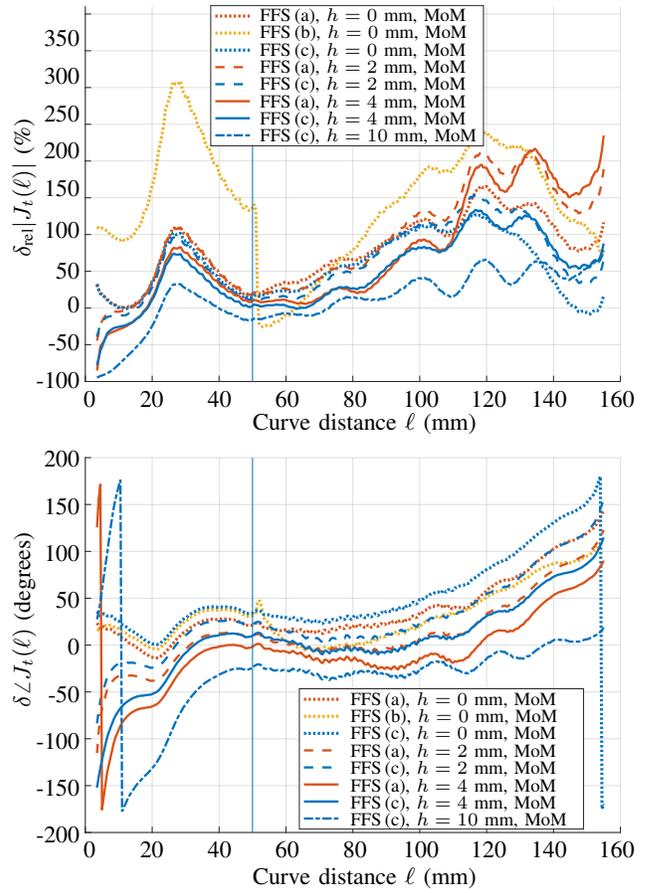
\begin{figure}[!t] 
	\centering
	\setlength{\unitlength}{1mm}
	\begin{picture}(80, 58)(0, 0)
		\put(0,2){
			\includegraphics[width=78mm, trim=35mm 98mm 40mm 102mm, clip] {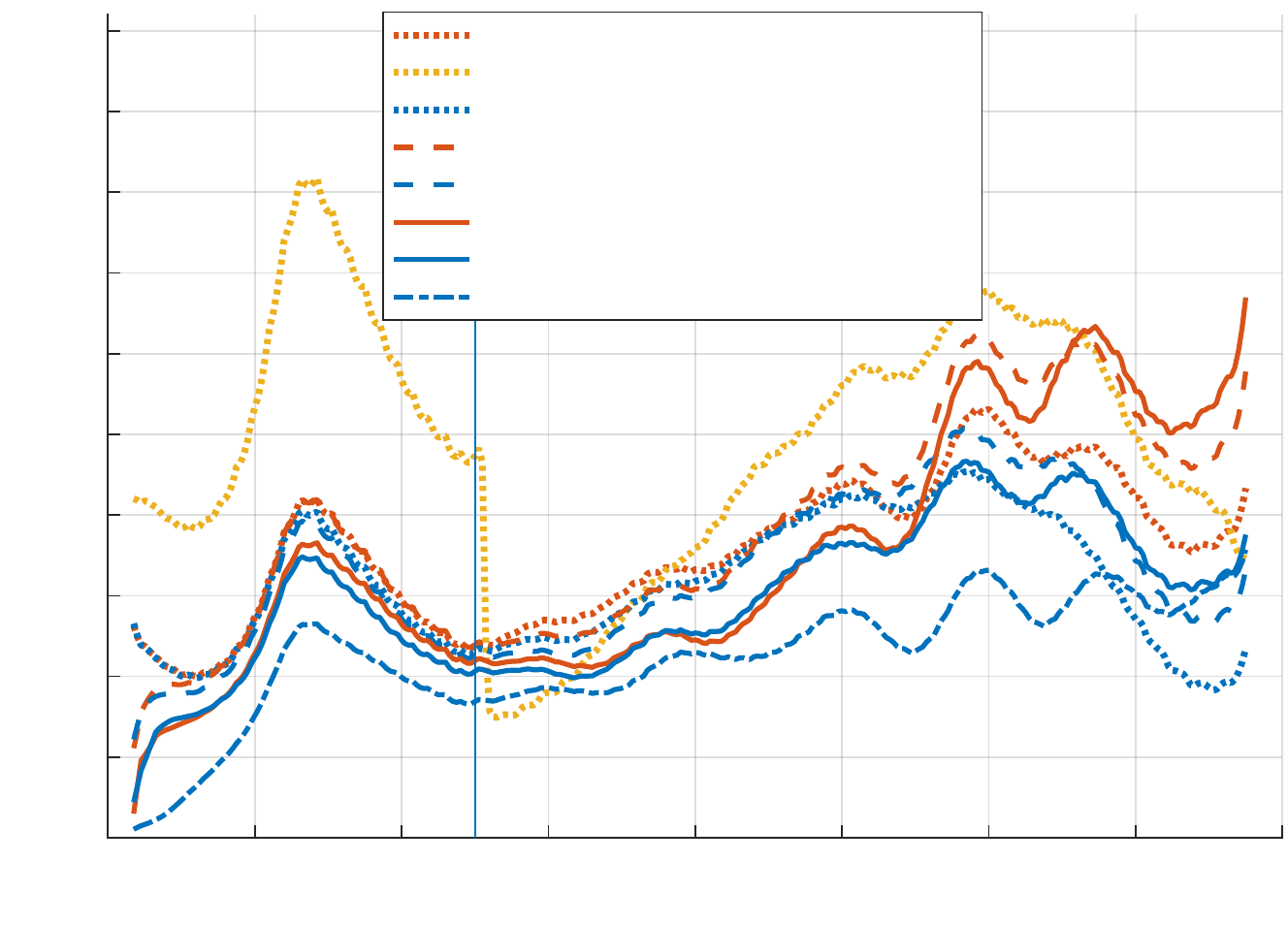}
		}
		\small 
		\put(30,1){\makebox(0,0)[lb]{\smash{Curve distance $\ell$ (mm)}}}%
		\put( 0.4,24){\rotatebox{90}{\makebox(0,0)[lb]{\smash{$\delta_\text{rel} |J_t(\ell)|$ (\%)}}}}%

		\put(7.5,4){\makebox(0,0)[lb]{\smash{0}}}%
   		\put(15, 4){\makebox(0,0)[lb]{\smash{20}}}%
   		\put(24, 4){\makebox(0,0)[lb]{\smash{40}}}%
   		\put(33, 4){\makebox(0,0)[lb]{\smash{60}}}%
   		\put(42, 4){\makebox(0,0)[lb]{\smash{80}}}%
   		\put(50, 4){\makebox(0,0)[lb]{\smash{100}}}%
   		\put(59, 4){\makebox(0,0)[lb]{\smash{120}}}%
   		\put(67, 4){\makebox(0,0)[lb]{\smash{140}}}%
   		\put(76, 4){\makebox(0,0)[lb]{\smash{160}}}%
	
   		\put( 7.0,52.0){\makebox(0,0)[rb]{\smash{350}}}%
   		\put( 7.0,47.0){\makebox(0,0)[rb]{\smash{300}}}%
   		\put( 7.0,42.0){\makebox(0,0)[rb]{\smash{250}}}%
   		\put( 7.0,37.0){\makebox(0,0)[rb]{\smash{200}}}%
   		\put( 7.0,32.0){\makebox(0,0)[rb]{\smash{150}}}%
   		\put( 7.0,26.5){\makebox(0,0)[rb]{\smash{100}}}%
   		\put( 7.0,21.0){\makebox(0,0)[rb]{\smash{50}}}%
   		\put( 7.0,16.5){\makebox(0,0)[rb]{\smash{0}}}%
   		\put( 7.0,11.5){\makebox(0,0)[rb]{\smash{-50}}}%
   		\put( 7.0, 6.5){\makebox(0,0)[rb]{\smash{-100}}}%
   		
		\scriptsize
		\scriptsize
		\put(31,55.25){\makebox(0,0)[lb]{\smash{FFS$\,$(a), $h=0$~mm, MoM}}}%
		\put(31,53.00){\makebox(0,0)[lb]{\smash{FFS$\,$(b), $h=0$~mm, MoM}}}%
		\put(31,50.75){\makebox(0,0)[lb]{\smash{FFS$\,$(c), $h=0$~mm, MoM}}}%
		\put(31,48.45){\makebox(0,0)[lb]{\smash{FFS$\,$(a), $h=2$~mm, MoM}}}%
		\put(31,46.15){\makebox(0,0)[lb]{\smash{FFS$\,$(c), $h=2$~mm, MoM}}}%
		\put(31,43.85){\makebox(0,0)[lb]{\smash{FFS$\,$(a), $h=4$~mm, MoM}}}%
		\put(31,41.55){\makebox(0,0)[lb]{\smash{FFS$\,$(c), $h=4$~mm, MoM}}}%
		\put(31,39.25){\makebox(0,0)[lb]{\smash{FFS$\,$(c), $h=10$~mm, MoM}}}%
	\end{picture}			
\\	
\vspace{2mm}
	\setlength{\unitlength}{1mm}
	\begin{picture}(80, 58)(0, 0)
		\put(0,2){
			\includegraphics[width=78mm, trim=35mm 98mm 40mm 102mm, clip] {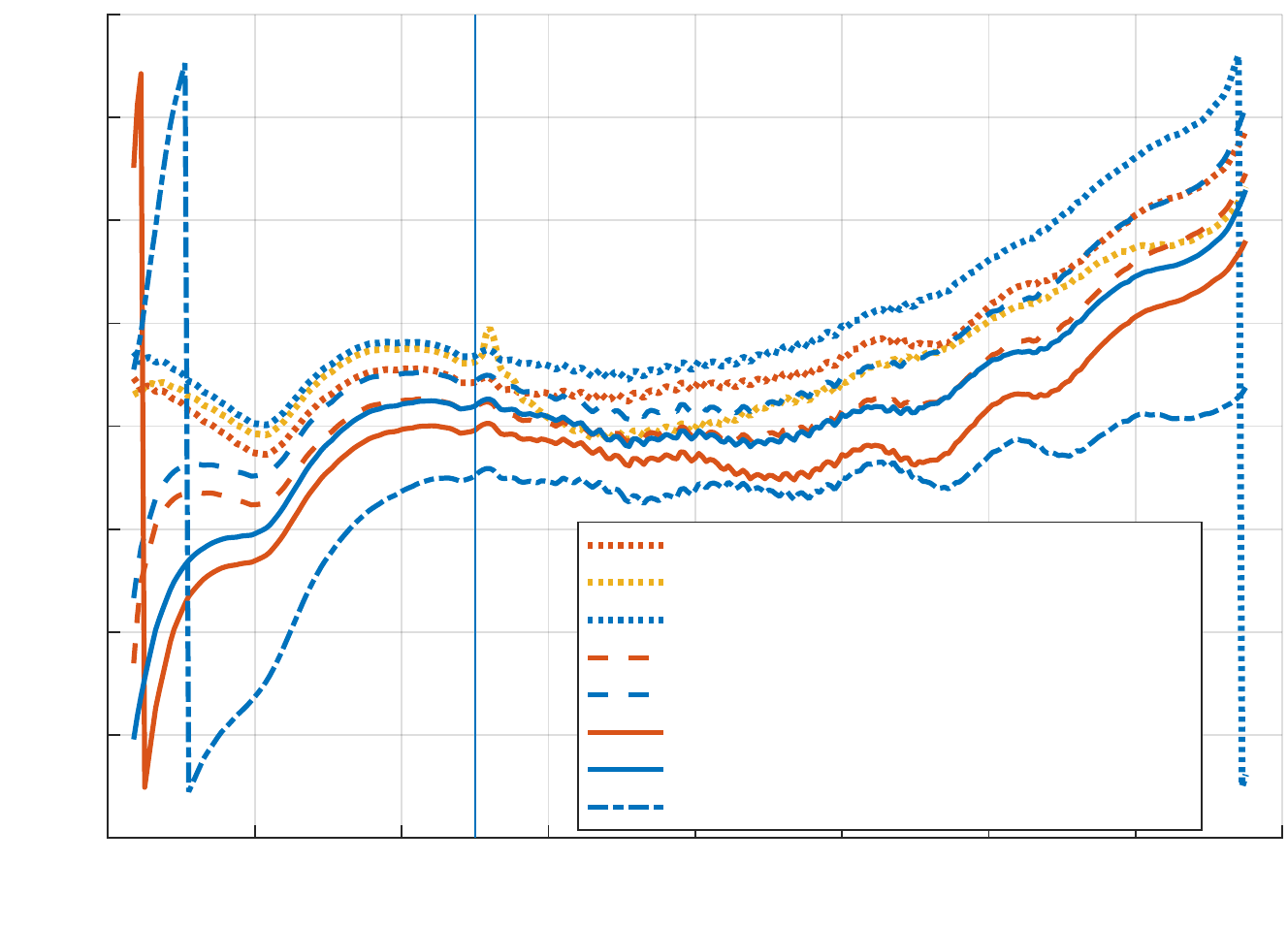}
		}
		\small 
		\put(30,0.5){\makebox(0,0)[lb]{\smash{Curve distance $\ell$ (mm)}}}%
		\put( 0.4,24){\rotatebox{90}{\makebox(0,0)[lb]{\smash{$\delta\protect\angle J_t(\ell)$ (degrees)}}}}%

		\put(7.5,4){\makebox(0,0)[lb]{\smash{0}}}%
   		\put(15, 4){\makebox(0,0)[lb]{\smash{20}}}%
   		\put(24, 4){\makebox(0,0)[lb]{\smash{40}}}%
   		\put(33, 4){\makebox(0,0)[lb]{\smash{60}}}%
   		\put(42, 4){\makebox(0,0)[lb]{\smash{80}}}%
   		\put(50, 4){\makebox(0,0)[lb]{\smash{100}}}%
   		\put(59, 4){\makebox(0,0)[lb]{\smash{120}}}%
   		\put(67, 4){\makebox(0,0)[lb]{\smash{140}}}%
   		\put(76, 4){\makebox(0,0)[lb]{\smash{160}}}%
	
   		\put( 7.0,56.0){\makebox(0,0)[rb]{\smash{200}}}%
   		\put( 7.0,49.75){\makebox(0,0)[rb]{\smash{150}}}%
   		\put( 7.0,43.5){\makebox(0,0)[rb]{\smash{100}}}%
   		\put( 7.0,37.25){\makebox(0,0)[rb]{\smash{50}}}%
   		\put( 7.0,31.00){\makebox(0,0)[rb]{\smash{0}}}%
   		\put( 7.0,24.75){\makebox(0,0)[rb]{\smash{-50}}}%
   		\put( 7.0,18.50){\makebox(0,0)[rb]{\smash{-100}}}%
   		\put( 7.0,12.25){\makebox(0,0)[rb]{\smash{-150}}}%
   		\put( 7.0, 6.00){\makebox(0,0)[rb]{\smash{-200}}}%
   		
		\scriptsize
		\put(43,24.25){\makebox(0,0)[lb]{\smash{FFS$\,$(a), $h=0$~mm, MoM}}}%
		\put(43,22.00){\makebox(0,0)[lb]{\smash{FFS$\,$(b), $h=0$~mm, MoM}}}%
		\put(43,19.75){\makebox(0,0)[lb]{\smash{FFS$\,$(c), $h=0$~mm, MoM}}}%
		\put(43,17.45){\makebox(0,0)[lb]{\smash{FFS$\,$(a), $h=2$~mm, MoM}}}%
		\put(43,15.15){\makebox(0,0)[lb]{\smash{FFS$\,$(c), $h=2$~mm, MoM}}}%
		\put(43,12.85){\makebox(0,0)[lb]{\smash{FFS$\,$(a), $h=4$~mm, MoM}}}%
		\put(43,10.55){\makebox(0,0)[lb]{\smash{FFS$\,$(c), $h=4$~mm, MoM}}}%
		\put(43,08.25){\makebox(0,0)[lb]{\smash{FFS$\,$(c), $h=10$~mm, MoM}}}%
  		\end{picture}	
	
	\caption{
	The relative current magnitude errors $\delta_\text{rel} |J_{t}(\ell)|$ (top) and current phase errors $\delta\protect\angle J_{t}(\ell)$ (bottom) evaluated along the curve in Fig.~\ref{fig:main_structure_with_curve}(b) using the far-field sources in Fig.~\ref{fig:conf_eq_farfield}, installed on height $h$ above the flat top  of $G$.
	}
	\label{fig:res_currents_H2_eq_farfield}
\end{figure}

\begin{figure}[!t] 
	\centering

		\setlength{\unitlength}{1mm}
		\begin{picture}(80, 58)(0, -2)
			\put(0,0){
				\includegraphics[width=78mm, trim=35mm 98mm 40mm 102mm, clip] {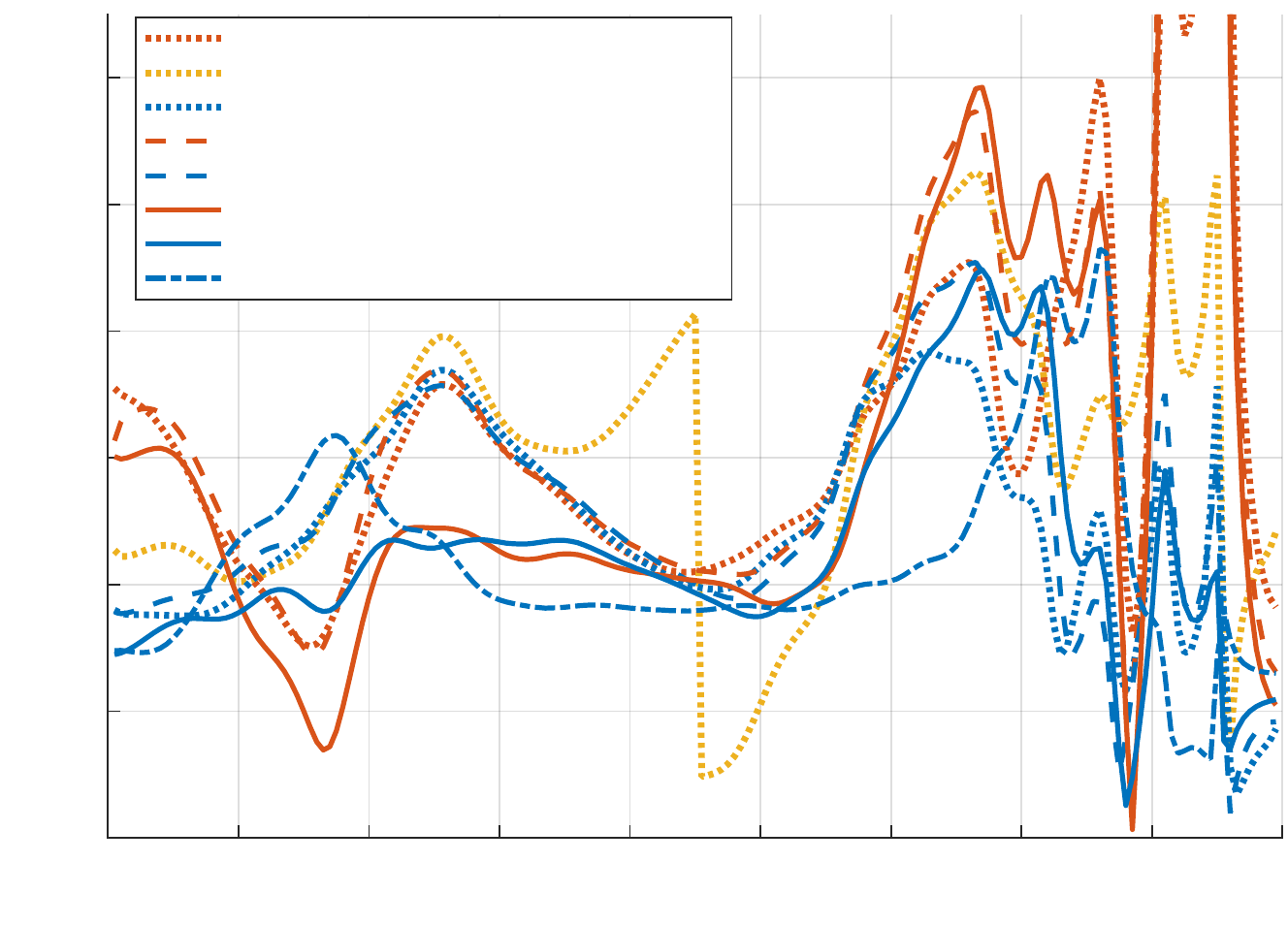}
			}
			
			\small 
		    \put(30,-2){\makebox(0,0)[lb]{\smash{Inclination angle $\theta$ (degrees)}}}%
		    \put( -0,15){\rotatebox{90}{\makebox(0,0)[lb]{\smash{$\delta_\text{rel} |E_\theta(90^\circ\!, \theta)|$ (\%)}}}}%

       		\put(7.5,2){\makebox(0,0)[lb]{\smash{0}}}%
    		\put(14, 2){\makebox(0,0)[lb]{\smash{20}}}%
    		\put(22, 2){\makebox(0,0)[lb]{\smash{40}}}%
    		\put(30, 2){\makebox(0,0)[lb]{\smash{60}}}%
    		\put(38, 2){\makebox(0,0)[lb]{\smash{80}}}%
    		\put(45, 2){\makebox(0,0)[lb]{\smash{100}}}%
    		\put(53, 2){\makebox(0,0)[lb]{\smash{120}}}%
    		\put(60, 2){\makebox(0,0)[lb]{\smash{140}}}%
    		\put(68, 2){\makebox(0,0)[lb]{\smash{160}}}%
    		\put(76, 2){\makebox(0,0)[lb]{\smash{180}}}%

    		\put( 7.0,51.2){\makebox(0,0)[rb]{\smash{200}}}%
    		\put( 7.0,43.5){\makebox(0,0)[rb]{\smash{150}}}%
    		\put( 7.0,35.75){\makebox(0,0)[rb]{\smash{100}}}%
    		\put( 7.0,28.4){\makebox(0,0)[rb]{\smash{50}}}%
    		\put( 7.0,19.75){\makebox(0,0)[rb]{\smash{0}}}%
    		\put( 7.0,12){\makebox(0,0)[rb]{\smash{-50}}}%
    		\put( 7.0, 5.0){\makebox(0,0)[rb]{\smash{-100}}}%

    		\scriptsize
		    \put(16,52.90){\makebox(0,0)[lb]{\smash{FFS$\,$(a), $h=0$~mm, MoM}}}%
			\put(16,50.79){\makebox(0,0)[lb]{\smash{FFS$\,$(b), $h=0$~mm, MoM}}}%
			\put(16,48.70){\makebox(0,0)[lb]{\smash{FFS$\,$(c), $h=0$~mm, MoM}}}%
		    \put(16,46.62){\makebox(0,0)[lb]{\smash{FFS$\,$(a), $h=2$~mm, MoM}}}%
			\put(16,44.54){\makebox(0,0)[lb]{\smash{FFS$\,$(c), $h=2$~mm, MoM}}}%
		    \put(16,42.46){\makebox(0,0)[lb]{\smash{FFS$\,$(a), $h=4$~mm, MoM}}}%
		    \put(16,40.38){\makebox(0,0)[lb]{\smash{FFS$\,$(c), $h=4$~mm, MoM}}}%
			\put(16,38.30){\makebox(0,0)[lb]{\smash{FFS$\,$(c), $h=10$~mm, MoM}}}%
   		\end{picture}	
	\caption{
	The relative installed far-field errors $\delta_\text{rel} |E_\theta(90^\circ, \theta)|$ using the far-field sources in Fig.~\ref{fig:conf_eq_farfield} installed on height $h$ above the flat top  of $G$. 
	}
	\label{fig:res_farfields_eq_farfield}

\end{figure}
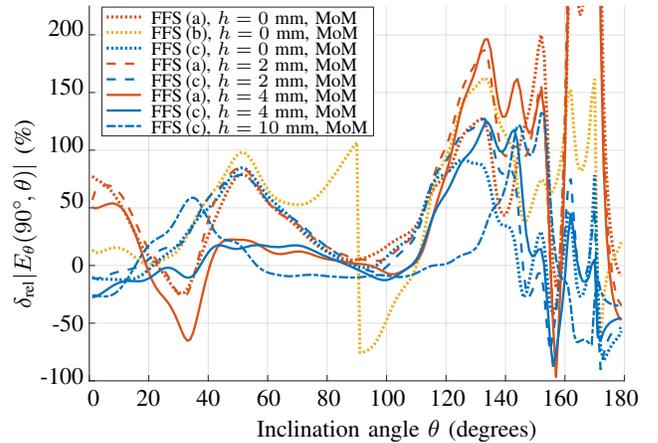

The accuracy of the far-field sources are evaluated with the current errors $\delta_\text{rel} |J_t(\ell)|$ and $\delta\angle J_t(\ell)$, according to \eqref{eq:measure_dJt_rel} and \eqref{eq:measure_dJt_phase}. 
These errors are depicted in Fig.~\ref{fig:res_currents_H2_eq_farfield} and listed as RMS errors in Table~\ref{tab:rms_ffs} for the investigated configurations.
The {far-field errors} $\delta_\text{rel} |E_\theta(\varphi,\theta)|$, according to \eqref{eq:measure_dE_rel}, are depicted in Fig.~\ref{fig:res_farfields_eq_farfield}. 

The ground plane in (c) is smaller than the flat surface of $G$. {Despite that, as depicted in Fig.~\ref{fig:res_eq_farfield}, far-field source (c) radiates less in the lower hemisphere}, as compared with (a) and (b). We note in Table~\ref{tab:rms_ffs} that the far-field error $\delta_\text{rel} |E_\theta(\varphi,\theta)|$ is smallest with (c), especially for $\theta > 90^\circ$. {This is somewhat surprising, since (a) captures the local geometry of the platform better.} However, the asymmetry of (c) makes it less attractive to use. 

If (b) is installed on a height $h > 0$ mm, there are no fields impinging on the platform, resulting in zero currents on the platform and also zero field for $\theta > 90^\circ$ (which is the reason for the omitted numbers in Table~\ref{tab:rms_ffs}). 
Hence, when generating a far-field source using an infinite ground plane, the resulting far-field source should be installed on the platform surface, i.e. $h=0$. For the other far-field sources, i.e. (a) and (c), it is hard to give any recommendations for the value of $h$.

{Compared with the estimated RMS uncertainty in the reference solutions in Section~\ref{sec:Results0} ($3.4\,$\% for the current magnitude and $3.1\,$\% for the installed far-field magnitudes), we see in Table~\ref{tab:rms_ffs} that the far-field sources increase the current magnitude uncertainty by a factor of $11$--$48$ 
and the  far-field magnitude uncertainty by a factor of $13$--$43$ 
(for $\theta \in (0, 180^\circ)$).}

Since the near-field behavior is not captured with far-field sources, it is expected that the surface currents are inaccurate. It is notable, however, that also the installed far-fields depicted in Fig.~\ref{fig:res_farfields_eq_farfield} are {more} inaccurate, as compared to using NFS.
It is also notable that the placement $h$ of the far-field source has such a strong impact, {see Fig.~\ref{fig:res_currents_H2_eq_farfield}--\ref{fig:res_farfields_eq_farfield}}. Its influence is in same order of magnitude as {the choice of} configuration. 

\begin{table}[!t] 
  \centering
  \caption{Root-mean-square errors using far-field source and MoM.\vspace{-1mm}} 
	\begin{tabular}{|ll||r|r|r|r|}
		\hline
		\multicolumn{2}{|l||}{
			 		Configu-
		} 
		& \multicolumn{4}{c|}{RMS (linear scale)}\\
	  	\cline{3-6}
		\multicolumn{2}{|l||}{
			ration
		}
		& \parbox[t]{13.8mm}{$\!\delta_\text{rel} |J_{t}(\ell)|$,
						   $\ell\!\in$
						   } 
		& \parbox[t]{14.1mm}{$\!\delta\angle J_{t}(\ell)$,
						   $\ell\!\in$} 
		& \parbox[t]{11.6mm}{$\!\delta_\text{rel} |E_{\theta}|$,
						   $\theta\!\in$} 
		& \parbox[t]{11.6mm}{$\!\delta_\text{rel} |E_{\theta}|$,
						   $\theta\!\in$} 
		\\
		\multicolumn{2}{|r||}{
			$h=$ 
		}
		& \parbox[t]{13.9mm}{$\!(3, 155]\,$mm} 
		& \parbox[t]{14.5mm}{$(3, 155]\,$mm} 
		& \parbox[t]{11.6mm}{$(0, 180^\circ\!)$} 
		& \parbox[t]{11.6mm}{$(0, 90^\circ\!)$} 
		 \\
	 	\hline

		FFS$\,$(a)
		&\multirow{3}{*}{ \rotatebox{90}{$0$ mm}}
		&  $89\,$\%   
		&  $52^\circ$ 
		& $133\,$\%   
		& $44\,$\%    
		\\
		FFS$\,$(b)
		&
		& $163\,$\% 
		&  $44^\circ$ &  $77\,$\% & $57\,$\% \\
		FFS$\,$(c)
		&
		&  $70\,$\% &  $69^\circ$ &  $46\,$\% & $41\,$\% \\
		\cline{1-2}
		
		FFS$\,$(a)
		&\multirow{3}{*}{ \rotatebox{90}{$2$ mm}}
		& $112\,$\% &  $40^\circ$ & $133\,$\% & $47\,$\% \\
		FFS$\,$(b)
		&
		&       --  & 		--  &  $74\,$\% & $27\,$\% \\
		FFS$\,$(c)
		&
		&  $79\,$\% &  $52^\circ$ &  $54\,$\% & $41\,$\% \\
		\cline{1-2}
		
		FFS$\,$(a)
		&\multirow{3}{*}{ \rotatebox{90}{$4$ mm}}
		& $105\,$\% &  $45^\circ$ & $127\,$\% & $30\,$\% \\
		FFS$\,$(b)
		&
		&       --  &         --  &  $74\,$\% & $27\,$\% \\
		FFS$\,$(c)
		&
		&  $66\,$\% &  $44^\circ$ &  $45\,$\% & $14\,$\% \\
		\cline{1-2}
		
		FFS$\,$(a)
		&\multirow{3}{*}{ \rotatebox{90}{$10$ mm}}
		&  $61\,$\% &  $61^\circ$ & $115\,$\% & $31\,$\% \\
		FFS$\,$(b)
		&
		&       --  &         --  &  $74\,$\% & $27\,$\% \\
		FFS$\,$(c)
		&
		&  $39\,$\% &  $58^\circ$ &  $40\,$\% & $25\,$\% \\
		\hline
	\end{tabular}
	\label{tab:rms_ffs}
\end{table}

\subsection{Results on the Impact of the Numerical Method}
\label{sec:Results3}

The above {discussed} results evaluate the accuracy of two different types of equivalent antenna representations. To determine how the estimated accuracy depends on the choice of numerical method, we use the best near-field configurations, see Fig.~\ref{fig:res_farfields_eq_surf_full_view}, and the best far-field source, {see} Fig.~\ref{fig:res_farfields_eq_farfield}, with different numerical methods. 
For near-field sources, we compare the accuracy of FIT with the accuracy of MoM and SBR. For the far-field sources we compare the accuracy of MoM with SBR. We do not consider FIT for far-field sources since it is not implemented in the current version of CST. 

\begin{figure}[!t] 
	\centering
		\setlength{\unitlength}{1mm}
		\begin{picture}(80, 58)(0, -2)
			\put(0,0){
				\includegraphics[width=78mm, trim=35mm 98mm 40mm 102mm, clip] {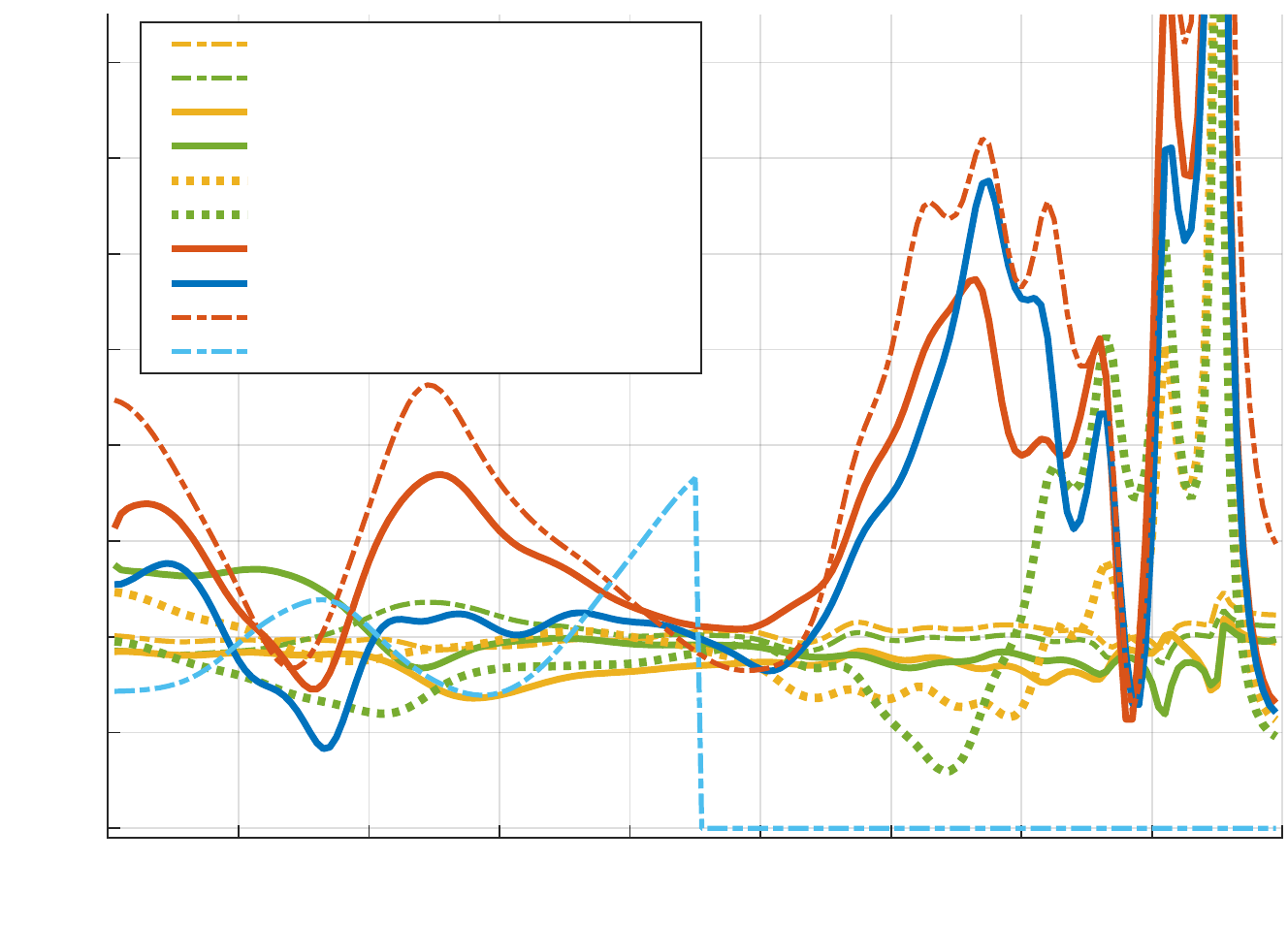}
			}
			
			\small 
		    \put(30,-2){\makebox(0,0)[lb]{\smash{Inclination angle $\theta$ (degrees)}}}%
		    \put( -0,15){\rotatebox{90}{\makebox(0,0)[lb]{\smash{$\delta_\text{rel} |E_\theta(90^\circ\!, \theta)|$ (\%)}}}}%

       		\put(7.5,2){\makebox(0,0)[lb]{\smash{0}}}%
    		\put(14, 2){\makebox(0,0)[lb]{\smash{20}}}%
    		\put(22, 2){\makebox(0,0)[lb]{\smash{40}}}%
    		\put(30, 2){\makebox(0,0)[lb]{\smash{60}}}%
    		\put(38, 2){\makebox(0,0)[lb]{\smash{80}}}%
    		\put(45, 2){\makebox(0,0)[lb]{\smash{100}}}%
    		\put(53, 2){\makebox(0,0)[lb]{\smash{120}}}%
    		\put(60, 2){\makebox(0,0)[lb]{\smash{140}}}%
    		\put(68, 2){\makebox(0,0)[lb]{\smash{160}}}%
    		\put(76, 2){\makebox(0,0)[lb]{\smash{180}}}%

    		\put( 7.0,51.25){\makebox(0,0)[rb]{\smash{300}}}%
    		\put( 7.0,45.50){\makebox(0,0)[rb]{\smash{250}}}%
    		\put( 7.0,39.75){\makebox(0,0)[rb]{\smash{200}}}%
    		\put( 7.0,33.75){\makebox(0,0)[rb]{\smash{150}}}%
    		\put( 7.0,28.00){\makebox(0,0)[rb]{\smash{100}}}%
    		\put( 7.0,22.50){\makebox(0,0)[rb]{\smash{50}}}%
    		\put( 7.0,17.00){\makebox(0,0)[rb]{\smash{0}}}%
    		\put( 7.0,11.25){\makebox(0,0)[rb]{\smash{-50}}}%
    		\put( 7.0, 4.75){\makebox(0,0)[rb]{\smash{-100}}}%

    		\scriptsize
		    \put(18,52.45){\makebox(0,0)[lb]{\smash{NFS$\,$(d), FIT}}}%
			\put(18,50.40){\makebox(0,0)[lb]{\smash{NFS$\,$(f), FIT}}}%
		    \put(18,48.35){\makebox(0,0)[lb]{\smash{NFS$\,$(d), MoM}}}%
			\put(18,46.30){\makebox(0,0)[lb]{\smash{NFS$\,$(f), MoM}}}%
		    \put(18,44.25){\makebox(0,0)[lb]{\smash{NFS$\,$(d), SBR}}}%
			\put(18,42.20){\makebox(0,0)[lb]{\smash{NFS$\,$(f), SBR}}}%
			\put(18,40.15){\makebox(0,0)[lb]{\smash{FFS$\,$(a), 2 mm, MoM}}}%
			\put(18,38.10){\makebox(0,0)[lb]{\smash{FFS$\,$(c), 4 mm, MoM}}}%
			\put(18,36.05){\makebox(0,0)[lb]{\smash{FFS$\,$(a), 2 mm, SBR}}}%
			\put(18,34){\makebox(0,0)[lb]{\smash{FFS$\,$(b), 4 mm, SBR}}}%
   		\end{picture}
	\caption{
	The relative installed far-field errors $\delta_\text{rel} |E_{\theta}(90^\circ,\theta)|$ using three different numerical methods; FIT, MoM, and SBR.
	}%
	\label{fig:res_inst_antenna_pattern_methods}
\end{figure}
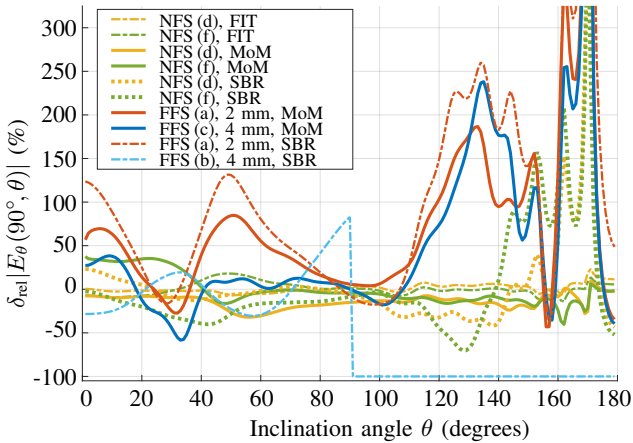

We use a subset of the equivalent sources from previous sections; three near-field sources; NFS~(b) with best $\delta_\text{rel} |E_\theta(90^\circ, \theta)|$, NFS~(d) with best $\delta_\text{rel} |J_{t}(\ell)|$, and NFS~(f) with best $\delta\angle J_{t}(\ell)$, and three far-field sources, FFS~(a) $h=2$~mm with best $\delta\angle J_{t}(\ell)$, FFS~(c) $h=4$~mm with best  $\delta_\text{} |E_{\theta}|$, and FFS~(c) $h=10$~mm with best $\delta_\text{rel} |J_{t}(\ell)|$.
The relative installed far-field errors $\delta_\text{rel} |E_\theta(90^\circ,\theta)|$, defined in \eqref{eq:measure_dE_rel}, from these equivalent sources are calculated with different numerical methods, FIT, MoM, and SBR. The resulting errors are depicted in Fig.~\ref{fig:res_inst_antenna_pattern_methods} and also listed as RMS errors in Table~\ref{tab:rms_methods}. 

\begin{table}[!t] 
  \centering
  \caption{Root-mean-square errors for different numerical methods and equivalent sources.\vspace{-1mm}}
	\begin{tabular}{|l||r|r|r|}
		\hline
		 	
		& 			 
		\multicolumn{3}{c|}{RMS (linear scale)}\\
	  	\cline{2-4}
		Configuration		
		& \multicolumn{3}{c|}{
		 		\parbox[t]{35mm}{
		 			$\delta_\text{rel} |E_{\theta}(90^\circ,\theta)|$, $\theta\!\in\!(0, 180^\circ\!)$
	 			} 
 			}
		\\
	  	\cline{2-4}
%
%
				   & \parbox[tc]{9mm}{FIT}
				   & \parbox[tc]{9mm}{MoM}
				   & \parbox[tc]{9mm}{SBR}
					 \\

%
%
  	 	\hline
		NFS$\,$(b) 			& $4.4\,$\% & $23\,$\% & $56\,$\% \\		
		NFS$\,$(d)          & $5.1\,$\% & $16\,$\% & $57\,$\% \\
		NFS$\,$(f)          & $7.3\,$\% & $18\,$\% & $65\,$\% 
							  \\
		\cline{1-1}
		FFS$\,$(a), $2\,$mm  & -- &  $133\,$\% &  $195\,$\%  \\
		FFS$\,$(b), $4\,$mm  & -- &  $74\,$\% &  $74\,$\%  \\
		FFS$\,$(c), $4\,$mm  & -- &  $45\,$\% &  $88\,$\%  \\
		FFS$\,$(c), $10\,$mm & -- &  $40\,$\% &  $91\,$\%  \\
		\hline
	\end{tabular}
	\label{tab:rms_methods}
\end{table}

We see in Table~\ref{tab:rms_methods} that, when using near-field sources, FIT performs significantly better than MoM and SBR. On average, {RMS errors are} $3$ times higher with MoM and $9$ times higher with SBR, as compared with FIT. 

When using far-field sources, MoM gives more accurate results than SBR, as seen in Table~\ref{tab:rms_methods}. None of the numerical methods give accurate results for $\theta > 90^\circ$ with far-field sources, see Fig.~\ref{fig:res_inst_antenna_pattern_methods}. With the combination of FFS~(b), $h > 0$ mm {and SBR}, there are no fields impinging on the platform, resulting in a zero field for $\theta > 90^\circ$. 
In Fig.~\ref{fig:res_inst_antenna_pattern_methods}, it is clear that the near-field sources are an order of magnitude more accurate than the far-field sources. Similar effects are observed in the currents as seen by comparing Table~\ref{tab:rms_eq_surf} with Table~\ref{tab:rms_ffs}.

\vspace{0pt}
%
\section{Discussion and Conclusions}
\label{sec:Conclusions}

Electromagnetic simulations of antennas installed on large platforms are challenging problems. The often complex antenna in combination with an electrically large platform leads to very high memory requirements and long simulation times. 
One way to reduce the complexity is to represent the antenna with an equivalent model that is more effective to use in simulations.

This paper presents one of the first accuracy studies of equivalent sources on platforms.
The presented work aims toward estimating the accuracy when using two different equivalent representations of antennas, near-field sources and far-field sources, installed on a simplified platform. Several different configurations has been considered, with respect to the approximation of the platform and geometrical parameters associated with the generation of the equivalent sources. 
{The determined deviation from the reference solution are presented for each of the examined configuration of the equivalent sources.}
The results can be used as {recommendations} for antenna designers and engineers {to choose} the most accurate equivalent source {configuration}. 


In agreement with previous knowledge, near-field sources perform significantly better than far-field sources for all configurations considered. The results with near-field sources are of the right order of magnitude for all configurations evaluated, while the errors with far-field sources are {unexpectedly} large for some configurations. 

We observe that near-field sources, in the {presence of a platform}, were comparably robust, with respect to location and size of the equivalent surface. 
The resulting RMS accuracies of the best cases evaluated are about $9\,$\% {and $4^\circ$} for the surface current {magnitude and phase, respectively,} and about $4\,$\% for the installed far-field magnitude. 



The accuracy of the far-field sources with respect to surface current phase is rather low. In our opinion, far-field sources should not be used when current phase information is required. In the best case investigated, the installed far-field RMS error on the magnitude is about $31\,$\% and for the current $23\,$\%. The installed height above the platform of the far-field source has a strong effect on the accuracy, which introduce an uncertainty in the use of far-field sources. One should bear in mind that a far-field source, {even though less accurate compared to a near-field source, is an efficient representation} to use in numerical calculations. If the {expected} accuracy is within requirements, far-field sources {can still be} an attractive representation.


For the implementations in CST Microwave Studio, the most accurate results in {these case-studies are obtained when using} near-field sources in combination with the full-wave solver FIT. With far-field sources, the accuracy is similar with MoM and SBR for directions within line-of-sight, while MoM performs better for non-line-of-sight directions. 



%
\vspace{0pt}
%

\FloatBarrier



%
\vspace{7pt}
\bibliographystyle{IEEEtran}
\bibliography{Library}

\begin{thebibliography}{10}
\providecommand{\url}[1]{#1}
\csname url@samestyle\endcsname
\providecommand{\newblock}{\relax}
\providecommand{\bibinfo}[2]{#2}
\providecommand{\BIBentrySTDinterwordspacing}{\spaceskip=0pt\relax}
\providecommand{\BIBentryALTinterwordstretchfactor}{4}
\providecommand{\BIBentryALTinterwordspacing}{\spaceskip=\fontdimen2\font plus
\BIBentryALTinterwordstretchfactor\fontdimen3\font minus
  \fontdimen4\font\relax}
\providecommand{\BIBforeignlanguage}[2]{{%
\expandafter\ifx\csname l@#1\endcsname\relax
\typeout{** WARNING: IEEEtran.bst: No hyphenation pattern has been}%
\typeout{** loaded for the language `#1'. Using the pattern for}%
\typeout{** the default language instead.}%
\else
\language=\csname l@#1\endcsname
\fi
#2}}
\providecommand{\BIBdecl}{\relax}
\BIBdecl

\bibitem{Macnamara2010}
T.~M. Macnamara, \emph{{Introduction to Antenna Placement and
  Installation}}.\hskip 1em plus 0.5em minus 0.4em\relax John Wiley \& Sons,
  2010.

\bibitem{EurAAP_WG4}
EurAAP, ``{EurAAP Working Group on Software (WG4)},'' 2016.

\bibitem{Vandenbosch2016}
G.~A.~E. Vandenbosch and F.~Mioc, ``{Bridging the simulations-measurements gap:
  State-of-the-art},'' in \emph{2016 10th Eur. Conf. Antennas Propag.}, 2016.

\bibitem{Vandenbosch2016a}
G.~A.~E. Vandenbosch, ``{Measurements and Simulations of the GSM Antenna},'' in
  \emph{2016 10th Eur. Conf. Antennas Propag.}, Davos, 2016.

\bibitem{RylanderEtAl2013}
T.~Rylander, P.~Ingelstr{\"{o}}m, and A.~Bondeson, \emph{{Computational
  Electromagnetics}}, 2nd~ed.\hskip 1em plus 0.5em minus 0.4em\relax
  Springer-Verlag New York, 2013, vol.~51.

\bibitem{ToselliWidlund2005}
A.~Toselli and O.~Widlund, \emph{{Domain Decomposition Methods – Algorithms
  and Theory}}, 1st~ed., ser. Springer Series in Computational
  Mathematics.\hskip 1em plus 0.5em minus 0.4em\relax Springer-Verlag Berlin
  Heidelberg, 2005.

\bibitem{Zhao2008}
K.~Zhao, V.~Rawat, and J.-F. Lee, ``{A Domain Decomposition Method for
  Electromagnetic Radiation and Scattering Analysis of Multi-Target
  Problems},'' \emph{IEEE Trans. Antennas Propag.}, vol.~56, no.~8, pp.
  2211--2221, aug 2008.

\bibitem{BeckerHansenhybrid}
A.~Becker and V.~Hansen, ``{A hybrid method combining the Time-Domain Method of
  Moments, the Time-Domain Uniform Theory of Diffraction and the FDTD},''
  \emph{Adv. Radio Sci.}, vol.~5, no.~6, pp. 107--113, 2007.

\bibitem{BarkaCaudrillier2007}
A.~Barka and P.~Caudrillier, ``{Domain Decomposition Method Based on
  Generalized Scattering Matrix for Installed Performance of Antennas on
  Aircraft},'' \emph{IEEE Trans. Antennas Propag.}, vol.~55, no. 6 II, pp.
  1833--1842, 2007.

\bibitem{Foged2015}
L.~J. Foged, L.~Scialacqua, F.~Saccardi, F.~Mioc, J.~L.~A. Quijano, and
  G.~Vecchi, ``{Antenna placement based on accurate measured source
  representation and numerical tools},'' in \emph{2015 IEEE Int. Symp. Antennas
  Propag. Usn. Natl. Radio Sci. Meet.}, 2015, pp. 1486--1487.

\bibitem{CSTMWS2016}
CST, ``{Microwave Studio},'' 2016.

\bibitem{HFSS2016}
Ansys, ``{HFSS},'' 2016.

\bibitem{FEKO2016}
Altair, ``{FEKO},'' 2016.

\bibitem{COMSOL2016}
COMSOL, ``{COMSOL Multiphysics},'' 2016.

\bibitem{Weiland2008}
T.~Weiland, M.~Timm, and I.~Munteanu, ``{A Practical Guide to 3-D
  Simulation},'' \emph{IEEE Microw. Mag.}, no. December, pp. 62--75, 2008.

\bibitem{Balanis1992}
C.~A. Balanis, ``{Antenna theory: a review},'' \emph{Proc. IEEE}, vol.~80,
  no.~1, pp. 7--23, 1992.

\bibitem{Balanis2005}
------, \emph{{Antenna Theory Analysis and Design}}, 3rd~ed.\hskip 1em plus
  0.5em minus 0.4em\relax John Wiley \& Sons, 2005.

\bibitem{Foged2014}
L.~J. Foged, L.~Scialacqua, F.~Saccardi, F.~Mioc, D.~Tallini, E.~Leroux,
  U.~Becker, J.~L. {Araque Quijano}, and G.~Vecchi, ``{Innovative
  representation of antenna measured sources for numerical simulations},'' in
  \emph{IEEE Antennas Propag. Soc. AP-S Int. Symp.}, 2014, pp. 2014--2015.

\bibitem{Wang2013}
H.~Wang, V.~Khilkevich, Y.~J. Zhang, and J.~Fan, ``{Estimating radio-frequency
  interference to an antenna due to near-field coupling using decomposition
  method based on reciprocity},'' \emph{IEEE Trans. Electromagn. Compat.},
  vol.~55, no.~6, pp. 1125--1131, 2013.

\bibitem{Li2016}
L.~Li, J.~Pan, C.~Hwang, and J.~Fan, ``{Radiation Noise Source Modeling and
  Application in Near-Field Coupling Estimation},'' \emph{IEEE Trans.
  Electromagn. Compat.}, vol.~58, no.~4, pp. 1314--1321, 2016.

\bibitem{Payet2014}
N.~Payet, M.~Darces, J.-l. Montmagnon, M.~H{\'{e}}lier, and F.~Jangal, ``{Near
  field to far field transformation by using equivalent sources in HF band},''
  in \emph{15th Int. Symp. Antenna Technol. Appl. Electromagn.}, 2012, pp.
  1--4.

\bibitem{Tanjong2010}
E.~Tanjong, ``{Modeling the Installed Performance of Antennas in a Ship Topside
  Environment},'' CST, Tech. Rep., 2010.

\bibitem{Love1901}
A.~E.~H. Love, ``{The Integration of the Equations of Propagation of Electric
  Waves},'' \emph{Philos. Trans. R. Soc. London. Ser. A, Contain. Pap. a Math.
  or Phys. Character}, vol. 197, pp. 1--45, 1901.

\bibitem{Schelkunoff1936}
S.~A. Schelkunoff, ``{Some Equivalence Theorems of Electromagnetics and Their
  Application to Radiation Problems},'' \emph{Bell Syst. Tech. J.}, vol.~15,
  no.~1, pp. 92--112, 1936.

\bibitem{StutzmanThiele1998}
W.~L. Stutzman and G.~A. Thiele, \emph{{Antenna Theory and Design}},
  2nd~ed.\hskip 1em plus 0.5em minus 0.4em\relax John Wiley \& Sons, 1998.

\bibitem{Balanis1997}
C.~A. Balanis, \emph{{Antenna Theory: Analysis and Design}}, 2nd~ed.\hskip 1em
  plus 0.5em minus 0.4em\relax John Wiley \& Sons, 1997.

\bibitem{Jackson1998}
J.~D. Jackson, \emph{{Classical Electrodynamics}}, 3rd~ed.\hskip 1em plus 0.5em
  minus 0.4em\relax John Wiley \& Sons, Inc., 1998.

\bibitem{Yaghjian1984}
A.~D. Yaghjian, ``{Equivalence of surface current and aperture field
  integrations for reflector antennas},'' \emph{IEEE Trans. Antennas Propag.},
  vol.~32, no.~12, pp. 1355--1358, 1984.

\bibitem{Gibson2008}
W.~C. Gibson, \emph{{The Method of Moments in Electromagnetics}}.\hskip 1em
  plus 0.5em minus 0.4em\relax Chapman \& Hall/CRC, 2008.

\bibitem{Makarov2002}
S.~N. Makarov, \emph{{Antenna and EM Modeling with MATLAB}}.\hskip 1em plus
  0.5em minus 0.4em\relax John Wiley \& Sons, 2002.

\bibitem{TafloveHagness2000}
A.~Taflove and S.~C. Hagness, \emph{{Computational Electrodynamics: The
  Finite-Difference Time-Domain Method}}, 2nd~ed.\hskip 1em plus 0.5em minus
  0.4em\relax Artech House, 2000.

\bibitem{Chew2001fast}
W.~C. Chew, J.-M. Jin, E.~Michielssen, and J.~Song, \emph{{Fast and Efficient
  Algorithms in Computational Electromagnetics}}, ser. Antennas and Propagation
  Library.\hskip 1em plus 0.5em minus 0.4em\relax Artech House, 2001.

\bibitem{LepvrierEtAl2014}
B.~L. Lepvrier, R.~Loison, R.~Gillard, L.~Patier, P.~Potier, and P.~Pouliguen,
  ``{Analysis of Surrounded Antennas Mounted on Large and Complex Structures
  Using a Hybrid Method},'' in \emph{8th Eur. Conf. Antennas Propag.}, 2014,
  pp. 2352--2355.

\bibitem{Malmstrom2016}
J.~Malmstr{\"{o}}m, H.~Frid, and B.~L.~G. Jonsson, ``{Approximate Methods to
  Determine the Isolation between Antennas on Vehicles},'' in \emph{2016 IEEE
  Antennas Propag. Soc. Int. Symp.}, 2016, pp. 131--132.

\bibitem{PathakWang1981}
P.~H. Pathak and N.~Wang, ``{Ray analysis of mutual coupling between antennas
  on a convex surface},'' \emph{IEEE Trans. Antennas Propag.}, vol.~29, no.~6,
  pp. 911--922, nov 1981.

\end{thebibliography}

%



\vspace{-5mm}
\begin{IEEEbiographynophoto}
{Johan Malmstr\"{o}m}
received the M.Sc. degree in electrical engineering from KTH Royal Institute of Technology, Stockholm, Sweden, in 2003. He has been in industry working with electromagnetic calculations since 2004 and is currently with Saab, Electronic Warfare division, Stockholm, Sweden. He is a graduate student at the Electromagnetic Engineering Laboratory, KTH Royal Institute of Technology, within the area of computational electromagnetics and electromagnetic compatibility. His research is supported by Saab.
\end{IEEEbiographynophoto}

\vspace{-5mm}
\begin{IEEEbiographynophoto}
{Henrik Holter}
received the M.Sc. degree in electrical engineering (with the distinction \textquotedblleft Best Graduate
of the Year\textquotedblright) and the Ph.D. degree in electromagnetic theory from KTH Royal Institute of Technology, Stockholm, Sweden, in 1996 and 2000, respectively. 
He is currently at Saab, Electronic Warfare division, Stockholm, Sweden, where he is head of strategy, research \& new technologies. 
He has authored over 35 journal and conference publications in the area of array antennas. Dr. Holter was the recipient of the IEEE Antennas and Propagation Society 2003 R.~W.~P. King Award.
\end{IEEEbiographynophoto}

\vspace{-4mm}
\begin{IEEEbiographynophoto}
{B.~L.~G.~Jonsson}
received the M.Sc. degree in engineering physics from Ume\aa{} University, Ume\aa{}, Sweden, in 1995, and the Ph.D. degree in electromagnetic theory in 2001 from KTH Royal Institute of Technology, Stockholm, Sweden. He was a postdoctoral fellow at University of Toronto, Canada and a Wissenschaftlicher Mitarbeiter (postdoc) at ETH Z\"{u}rich, Switzerland. Since 2006 he is with the Electromagnetic Engineering Laboratory at KTH where he is professor since 2015. His research interests include electromagnetic theory in a wide sense, including scattering, antenna theory and non-linear dynamics. 
\end{IEEEbiographynophoto}





\vspace{00mm}

\enlargethispage{-00mm}
\clearpage

\end{document}